\documentclass[journal=jctcce,manuscript=article]{achemso}

\usepackage[table]{xcolor}
\usepackage{amsmath}
\usepackage{amssymb}
\usepackage{amsfonts}
\usepackage{bm}
\usepackage{bbm}
\usepackage{braket}
\usepackage{booktabs}
\usepackage{mathtools}
\usepackage[Symbol]{upgreek}
\usepackage{hyperref}
\usepackage{microtype}
\usepackage{tikz}
\hypersetup{
    colorlinks=true,
    linkcolor=blue,
    citecolor=red,
    filecolor=magenta,      
    urlcolor=cyan
}
\usepackage{placeins}

\newcommand\nnfootnote[1]{%
  \begin{NoHyper}
  \renewcommand\thefootnote{}\footnote{#1}%
  \addtocounter{footnote}{-1}%
  \end{NoHyper}
}

\newcommand\identity{1\kern-0.25em\text{l}}
\definecolor{SchoolColor}{rgb}{0.0,0.470588,0.5803922} 

\title{Renormalized Internally-Contracted Multireference Coupled Cluster with Perturbative Triples}

\setkeys{acs}{ maxauthors = 0 }

\author{Robin Feldmann}
\affiliation{ETH Z{\"u}rich, Department of Chemistry and Applied Biosciences, Vladimir-Prelog-Weg 2, 8093 Z{\"u}rich, Switzerland}
\author{Markus Reiher}
\email{mreiher@ethz.ch}
\affiliation{ETH Z{\"u}rich, Department of Chemistry and Applied Biosciences, Vladimir-Prelog-Weg 2, 8093 Z{\"u}rich, Switzerland}

\date{August 19, 2024}

\usepackage{array}
\newcolumntype{R}[1]{>{\raggedleft\let\newline\\\arraybackslash\hspace{0pt}}m{#1}}

\begin{document}

\begin{abstract}
In this work, we combine the many-body formulation of the internally contracted multireference coupled cluster (ic-MRCC) method with Evangelista's multireference formulation of the driven similarity renormalization group (DSRG).
The DSRG method can be viewed as a unitary multireference coupled cluster theory, which renormalizes the amplitudes based on a flow equation approach to eliminate numerical instabilities. 
We extend this approach by demonstrating that the unitary flow equation approach can be adapted for nonunitary transformations, rationalizing the renormalization of ic-MRCC amplitudes. 
We denote the new approach, the renormalized ic-MRCC (ric-MRCC) method.
To achieve high accuracy with a reasonable computational cost, we introduce a new approximation to the Baker--Campbell--Hausdorff expansion. We fully consider the linear commutator while approximating the quadratic commutator, for which we neglect specific contractions involving amplitudes with active indices. Moreover, we introduce approximate perturbative triples to obtain the ric-MRCCSD[T] method.
We demonstrate the accuracy of our approaches in comparison to advanced multireference methods for the potential energy curves of H$_8$, F$_2$, H$_2$O, N$_2$, and Cr$_2$. 
Additionally, we show that ric-MRCCSD and ric-MRCSSD[T] match the accuracy of CCSD(T) for evaluating spectroscopic constants and of full configuration interaction energies for a set of small molecules.
\end{abstract}


\maketitle
\nnfootnote{
$^*$ Corresponding author.
}

\section{Introduction}

Approximate electronic structure models that construct the wave function from a single reference determinant, such as single-reference coupled cluster (SRCC), will fail if strong correlation in the full configuration interaction (FCI) wave function causes multiple determinants to have significant weights.\cite{Helgaker2014_Molecular,Bartlett2009_Book}
Methods targeting strong correlation typically start from the complete active space (CAS) concept,\cite{Roos1980_CASSCF,Ruedenberg1982_FORS,Roos1987_CompleteActive} which aims to calculate the FCI wave function within a limited subspace and are denoted as CASCI models. However, exact CASCI calculations are constrained by the exponential scaling with the number of orbitals, limiting the size of the CAS. Hence, significant research efforts\cite{Eriksen2020_BenzeneGroundState} have focused on developing efficient approximate FCI methods to overcome this limitation. These include selected CI\cite{Malrieu1973_SelectedCi1,Malrieu1983_SelectedCi2}, heat-bath CI,\cite{Holmes2016_HBCI1,Smith2017_HBCI2} density matrix renormalization group (DMRG) \cite{White1992_DMRG1,white1993_DMRG2,Chan2008_Review1,Chan2009_Review2,Marti2010_Review-DMRG,Schollwock2011_MPSReview,Chan2011d_DMRG3,Wouters2014_ReviewDMRG,Kurashige2014_ReviewDMRG,Olivares2015_DMRGInPractice,Szalay2015_TensorReview,Yanai2015_DMRGReview,Baiardi2020_Perspective}, FCI quantum Monte Carlo\cite{Booth2009_FCIQMC1,Cleland2011_FCIQMC2}, and many-body expanded FCI\cite{Eriksen2017_MbeFci} methods. 

With these approaches, it has become possible to address large active spaces, but for quantitative accuracy, correlations from orbitals outside the active space must be included. A conceptually straightforward way to account for this dynamic correlation is achieved by employing CC amplitudes obtained from CASCI calculations in an SRCC formalism. This approach was pioneered by Oliphant, Adamowicz, and Piecuch.\cite{Oliphant1991_ExternallyCorrected0,Oliphant1992_ExternallyCorrected1,Piecuch1993_SplitAmplitude1,Piecuch1994_SplitAmplitude2_H8,Piecuch1998_Review} Especially the split-amplitude ansatz\cite{Piecuch1993_SplitAmplitude1,Piecuch1994_SplitAmplitude2_H8,Piecuch1994_SplitAmplitude3_H8,Piecuch1995_SplitAmplitude4_H2O} paved the way for practical CC methods that can account for a small number of strongly correlated electrons. Particularly well-explored multireference-driven SRCC methods are externally-corrected CC,\cite{Paldus1997_ExternallyCorrected1,Paldus1998_ExternallyCorrected2,Paldus1994_ExternallyCorrected3,Paldus1994_ExternallyCorrected4,Paldus1997_ExternallyCorrected5,Stolarczyk1994_ExternallyCorrected} tailored CC (TCC)\cite{Kinoshita2005_TailoredCC1,Hino2006_TailoredCC2,Veis2016_TCC-DMRG,Faulstich2019_TCCAnalysis,Pittner2019_LocalTCC,Pittner2020_LocalTCC,Faulstich2019_TCCNumerical,Kats2020_FCIQMC_TDCC1,Brandejs2020_RelativisticTCC,Boguslawski2021_TCC,Kats2022_FCIQMC_TDCC2,Morchen2022_TCC}, CAS-CC\cite{Adamowicz2000_cascc1,Ivanov2000_cascc2,Lyakh2005_cascc3}, method-of-moments CC,\cite{piecuch2004,kowalski2000,kowalski2000a} and the CC(P;Q) ansatz\cite{shen2012,shen2012a,shen2012b,bauman2017,deustua2017}. 
However, all of these approaches still construct the excitations systematically from a single determinant, which may have detrimental effects on the accuracy in strongly correlated systems.\cite{Morchen2022_TCC} 
Recently, we have investigated whether externally corrected or TCC methods can be systematically improved,\cite{feldmann2024complete} but concluded that the bias inherent to the single-reference approach is difficult to eliminate.
Therefore, in this work, we shift our focus to genuine multireference methods, constructing excitations from a multideterminantal reference.

The most widely used multireference methods are multireference perturbation theory (MRPT) methods, and, in particular, CAS second-order\cite{Andersson1992_CASPT2} perturbation theory (CASPT2) and \textit{N}-electron valence state second-order perturbation theory (NEVPT2)\cite{Angeli2001_NelectronValence}. However, the accuracy of MRPT methods falls far short of the desired accuracy achievable with CC methods in the single reference regime.
To achieve a higher accuracy than MRPT, multireference CI (MRCI) approaches\cite{Buenker1974_SelectedMRCI1,Meyer1977_icMRCI1,Shavitt1977_CI,Siegbahn1980_icMRCI2,Werner1982_icMRCI3,Buenker1983_SelectedMRCI2,Werner1988_icMRCI,Shavitt1998_ucMRCI,Shamasundar2011_NewIcMRCI,Szalay2012_MrciReview,Lischka2018_MrCiAndMrCcReview} can be applied, which include all single- and double-excitations from the CAS wave function. 
MRCI methods with approximate size consistency corrections\cite{Szalay2012_MulticonfigurationSelfConsistent} deliver high accuracy over entire potential energy curves.
However, they are computationally expensive because they either involve diagonalizing the Hamiltonian in an exponentially growing space or they require high-order reduced density matrices (RDMs) to be evaluated. 

Consequently, much work has been directed toward developing genuine size-consistent multireference coupled cluster (MRCC) methods in the past decades.
Yet, to this point, there is no generally accepted MRCC method that serves as a proper alternative to CCSD(T) in multireference scenarios. In the following, we summarize past developments in the field and elaborate on the issues that prohibit most MRCC methods from general applicability. For more in-depth discussions, we refer to Refs.~\citenum{Jeziorski2010_MultireferenceCoupledcluster,Lyakh2012_MultireferenceNature,Kohn2013_StatespecificMultireference,Evangelista2018_PerspectiveMultireference,Lischka2018_MrCiAndMrCcReview}.

Genuine MRCC methods can be categorized according to the application of the exponential operator. The operator can be applied to each individual determinant within the reference space or to the entire multireference wave function.
The former approach is known as the Jeziorski--Monkhorst (JM) ansatz\cite{Jeziorski1981_JMMMRCC}, which serves as a basis for both multi-state (state universal) and state-specific MRCC formulations.
Because the multi-state approaches are susceptible to the intruder state problem,\cite{Paldus1993_ApplicationHilbertspace,Kowalski2000_CompleteSet} more research has focused on the state-specific variant.
One of the pioneering efforts in this direction was undertaken by Laidig and Bartlett\cite{Laidig1984_MultireferenceCoupledcluster}, and
several other variants followed, among those are Mukherjee MRCC (Mk-MRCC) \cite{Mahapatra1998_MkMRCC1,Mahapatra1999_MkMRCC2,Maitra2012_MkMRCC3}, Brillouin--Wigner MRCC \cite{Mavsik1998_BWMRCC}, and the single-root MRCC\cite{Mahapatra2010_SingleRootMrCC}. 
Especially the Mk-MRCC method was extensively studied\cite{Evangelista2006_HighorderExcitations,Evangelista2007_CouplingTerm,Bhaskaran-Nair2010_MultireferenceMukherjee,Bhaskaran-Nair2011_MultireferenceStatespecific,Datta2011_SpinfreeAnalogue}, yet, it suffers from several problems\cite{Lyakh2012_MultireferenceNature}, such as convergence difficulties and the lack of invariance with respect to active orbital rotations.\cite{Evangelista2010_InsightsOrbital,Kong2010_OrbitalInvariance}
Some issues stemming from the fact that excitations applied do different determinants in the reference lead to the same excited determinants (multiple parentage problem) could be resolved by Hanrath\cite{Hanrath2005_MRexpT,Hanrath2006_MRexpT}, but the lack of orbital invariance persists for all JM-based methods. Moreover, the size of the active space is severely limited since the computational scaling is proportional to the number of determinants in the active space. 

In this work, we explore the internally contracted MRCC (ic-MRCC) method, in which the exponential operator is applied to a multideterminantal reference as a whole. 
This approach offers two advantages: the number of free parameters does not increase factorially with the size of the active space, and the ansatz remains invariant with respect to active orbital rotations\cite{Evangelista2011_OrbitalinvariantInternally}. However, its practical implementation was hindered for many years due to the complexity of the resulting equations.
The internally contracted ansatz was initially proposed by {\v C}{\'i}{\v z}ek\cite{Cizek1969_UseCluster} and was followed by Mukherjee\cite{Mukherjee1975_CorrelationProblem,Mukherjee1975_NonperturbativeOpenshell,Mukherjee1995_CoupledCluster,Mahapatra1998_StateSpecificMultiReference} and Banerjee and Simons.\cite{Banerjee1981_CoupledclusterMethod,Banerjee1982_ApplicationsMulticonfigurational,Banerjee1983_TestMulticonfigurational}. 
A significant development of ic-MRCC theories was the introduction of the generalized normal ordering by Mukherjee and Kutzelnigg,\cite{Mukherjee1997_NormalOrdering,Kutzelnigg1997_NormalOrdering} which extends the standard normal ordering to allow for excitation operators to act directly on a multideterminantal vacuum. 
However, due to the complexity of the ic-MRCC equations, the first implementations of the ic-MRCC method with singles and doubles (ic-MRCCSD) without approximations came later by
Evangelista and Gauss,\cite{Evangelista2011_OrbitalinvariantInternally} and by Hanauer and K{\"o}hn\cite{Hanauer2011_PilotApplications}. Hanauer and K{\"o}hn introduced a procedure to eliminate the multiple-parentage problem, where redundant excitations can be numerically eliminated with a singular value decomposition.
These pilot applications showed that ic-MRCC delivers remarkable accuracy while being size extensive and size consistent when operators are normal-ordered with respect to the generalized vacuum.\cite{Hanauer2012_CommunicationRestoring} 

Therefore, the ic-MRCC theory fulfills the essential criteria for MRCC methods: polynomial scaling relative to the active space size, invariance to active-orbital rotations, and both size extensivity and size consistency. 
Following the initial work, several extensions of the ic-MRCC method were proposed, such as a perturbative triples correction in line with the single-reference (T) correction\cite{Hanauer2012_PerturbativeTreatment}, explicit correlation\cite{Liu2013_ExplicitlyCorrelated}, and the calculation of excited states\cite{Samanta2014_ExcitedStates,Aoto2016_InternallyContracted}.
In parallel to these extensions, Nooijen, Datta, and Kong introduced the many-body\cite{Nooijen1996_GeneralSpin} approach to ic-MRCC with the partially internally-contracted  MRCCSD method\cite{Datta2011_StatespecificPartially} and the MR equation-of-motion CC.\cite{Datta2012_MultireferenceEquationofmotion,Demel2013_AdditionalGlobal,Nooijen2014_CommunicationMultireference} 
In this formulation, off-diagonal elements of the similarity transformed Hamiltonian serve as residuals for the amplitudes in contrast to the projected CC residuals.\cite{Datta2011_StatespecificPartially} This completely eliminates the multiple-parentage problem, but the equations may become numerically unstable when the active space contains redundant orbitals. The difference between the many-body and the projected residuals is discussed in detail in Lechner's PhD thesis.\cite{lechner2023_icMRCC-Thesis} 

In addition to standard ic-MRCC theory, a unitary version has been developed,\cite{Hoffmann1988_UnitaryMulticonfigurational,Chen2012_OrbitallyInvariant} introducing a distinct class of methods.
These unitary ic-MRCC methods share similarities with similarity renormalization group (SRG)-based approaches.
To recognize the similarities, we recall that CC theory can be viewed as an effective Hamiltonian theory, where the off-diagonal occupied-virtual elements are fully decoupled for optimized amplitudes. 
Similarly, the SRG employs a unitary flow to decouple the off-diagonal elements of the Hamiltonian symmetrically. 
Contracted Schrödinger equation-based methods\cite{Nakatsuji1976_EquationDirect,Cohen1976_HierarchyEquations,Colmenero1993_ApproximatingQorder,Colmenero1993_ApproximatingQordera,Nakatsuji1996_DirectDetermination,Yasuda1997_DirectDetermination,Mazziotti1998_ContractedSchr,Mazziotti1998_ApproximateSolution} are also related to unitary transformation theories. Specifically, the Heisenberg representation of the anti-hermitian contracted Schrödinger equation\cite{Mazziotti2006_AntiHermitianContracted,Mazziotti2007_AntiHermitianPart} employs a unitary flow equation for the Hamiltonian.

Initially formulated by Wegner,\cite{Wegner1994_FlowequationsHamiltoniansa} and by Glazek and Wilson\cite{Glazek1993_RenormalizationHamiltonians,Glazek1994_PerturbativeRenormalization}, the SRG has been extensively applied in various areas of physics.\cite{Kehrein2006_FlowEquation} Notably, White was the first to adapt SRG for chemical Hamiltonians through his canonical transformation (CT) theory.\cite{White2002_NumericalCanonical}
Yanai and Chan further advanced White's CT theory by employing a cumulant expansion, also utilized in contracted Schrödinger equation-based methods.\cite{Yanai2006_CanonicalTransformation,Yanai2007_CanonicalTransformation,Nakatsuji1996_DirectDetermination,Yasuda1997_DirectDetermination,Mazziotti1999_ComparisonContracted,Mazziotti2000_CompleteReconstruction,Mukherjee2001_IrreducibleBrillouin} 
In particular, the non-truncating Baker--Campbell--Hausdorff (BCH) expansion is approximated with the cumulant expansion, such that higher-order commutators can be approximated by nested evaluations of the one- and two-body contribution of the linear commutator.\cite{Yanai2006_CanonicalTransformation}
The approximation to the transformed Hamiltonian is called the recursive single commutator approximation.\cite{Yanai2006_CanonicalTransformation} 
This approximation was later studied by Evangelista and Gauss\cite{Evangelista2012_ApproximationSimilaritytransformed}, who showed that the approximation may introduce substantial errors due to mismatched prefactors for higher-order commutators. To be more concise, the largest error is introduced in the calculation of the double commutator of the doubles amplitudes with the fluctuation potential where an additional factor of $\frac{1}{2}$ is introduced. 
Nonetheless, the numerical studies carried out with CT theory demonstrated accurate results, also for strongly correlated systems.\cite{Yanai2006_CanonicalTransformation,Yanai2007_CanonicalTransformation,Neuscamman2009_QuadraticCanonical,Neuscamman2010_ReviewCanonical} 
Still, CT theory shows some deficiencies similar to the standard formulation of ic-MRCC, such as a dependence on the threshold for discarding redundancies, which also may cause numerical instability. Additionally, the theory scales with the ninth power in the number of active orbitals, which severely limits the size of the active space. 

The in-medium SRG (IM-SRG) method has recently attracted attention in nuclear structure theory as an effective extension of the SRG to the normal-ordered $n$-body second quantized spaces and has also been adapted to a multireference theory.\cite{Tsukiyama2011_InMediumSimilarity,Tsukiyama2012_InmediumSimilarity,Hergert2016_InMediumSimilaritya,Gebrerufael2017_InitioDescription}
The IM-SRG method inspired Evangelista's development of the driven SRG (DSRG) approach.\cite{Evangelista2014_DSRG1} Instead of directly solving the flow equation as in the IM-SRG, the DSRG optimizes amplitude equations analogously to CC theory.\cite{Evangelista2014_DSRG1,Evangelista2019_DSRG-Rev} 
Initially, the DSRG was formulated as a single-reference unitary coupled-cluster-like theory, in which the BCH expansion is evaluated with a recursive single commutator approximation, and the amplitude equations are renormalized by a transformation derived from the SRG. 
Soon after, Li and Evangelista presented a DSRG-MRPT\cite{Evangelista2015_MRDSRG,Evangelista2017_DSRG_PT3,Evangelista2020_ldsrg3}, and the iterative multireference method MR-LDSRG(2),\cite{Evangelista2016_RelaxedDSRG} where `L' denotes `linearized' since only the linear commutator is evaluated and `(2)' signals that only up to two-body operators are retained in the recursive linear commutator expansion.
In MR-LDSRG(2), the singles and doubles many-body residuals are optimized, and the renormalization eliminates the problem that internally contracted theories are numerically unstable when the active space contains nearly redundant orbitals,\cite{Yanai2007_CanonicalTransformation,Datta2011_StatespecificPartially} i.e., orbitals with small natural occupation numbers.
Nonetheless, two issues remain: the recursive commutator expansion may introduce errors,\cite{Evangelista2012_ApproximationSimilaritytransformed} and the energy depends on a free parameter introduced by renormalization. 

Our work builds on the MR-LDSRG(2) approach. We address the first issue and connect the nonunitary ic-MRCC theory with the unitary MR-LDSRG(2) by showing that the arguments for a unitary flow equation, in line with the IM-SRG,\cite{Wegner1994_FlowequationsHamiltoniansa,Kehrein2006_FlowEquation,Tsukiyama2011_InMediumSimilarity} can be extended to a nonunitary flow. Solving the flow equations does not introduce a free parameter but bears other challenges, such as integrating the differential equations.\cite{Morris2015_MagnusExpansion}
We transfer Evangelista's renormalization procedure to the nonunitary case, which we call renormalized ic-MRCC (ric-MRCC).
We demonstrate several advantages.
For example, the DSRG relies on the recursive commutator expansion, which can become computationally expensive.
We found that for MR-LDSRG(2), up to ten nested commutators are required to converge the transformation in every iteration of the amplitude optimization. 
By contrast, nonunitary ic-MRCCSD was shown to be highly accurate even if only the linear and double commutators were evaluated.\cite{Evangelista2011_OrbitalinvariantInternally} Nonetheless, an exact evaluation of the double commutator still remains prohibitively costly.
Guided by Hanauer and Köhn's observations\cite{Hanauer2011_PilotApplications,Hanauer2012_PerturbativeTreatment} that amplitudes with multiple active indices tend to be redundant, our strategy simplifies the second commutator by excluding contributions from equations involving these amplitudes. 
Moreover, the faster convergence of the similarity transformation and the flexibility to neglect some amplitudes allow us to introduce a computationally efficient perturbative triples correction.\cite{Hanauer2012_PerturbativeTreatment}

This work is organized as follows: Sec.~\ref{sec:theory_mrcc} introduces the generalized normal ordering, nonunitary SRG, and ric-MRCC. We then detail our approach to approximating the double commutator and discuss the perturbative triples correction. In Sec.~\ref{sec:comp_details_mrcc}, we outline the computational methods used to generate the results discussed in Sec.~\ref{sec:results_mrcc}. The latter includes an analysis of non-parallelity errors (NPEs) in the potential energy curves of several strongly correlated systems, comparing spectroscopic constants with CCSD(T) results and experimental data, and comparing total electronic energies for a test set of small molecules.

\section{Theory}
\label{sec:theory_mrcc}

\subsection{Generalized normal ordering}

We briefly review the generalized normal ordering\cite{Mukherjee1997_NormalOrdering,Kutzelnigg1997_NormalOrdering,Kong2010_AlgebraicProof}.
We define a general reference state $|\Psi_0\rangle$ given as a CASCI wave function
\begin{equation}
    |\Psi_0\rangle = \sum_{\mu=1}^d c_\mu |\Phi_\mu\rangle.
\label{eq:ref}
\end{equation}
The CASCI wave function is generated by systematically constructing all excitations of $N_A$ electrons within a subset of $A$ spin orbitals, $\mathbb{A}=\{\phi_u\}$. It is given in terms of spatial orbitals as CAS($N_A,A/2$).  We note here that the reference does not have to be an exact CASCI wave function, but it can also be a matrix product state\cite{Schollwock2011_MPSReview}, or any other approximate wave function for which reduced density matrices (RDMs) can be evaluated.\cite{Holmes2016_HBCI1,Booth2009_FCIQMC1}
The set of $C$ core orbitals $\mathbb{C}=\{\phi_m\}$ are all singly occupied in the CAS wave function, and the set of $V$ virtual orbitals $\mathbb{V}=\{\phi_e\}$ are unoccupied. The union of the sets $\mathbb{C}$, $\mathbb{A}$, and $\mathbb{V}$ is the set of all $N$ spin orbitals, $\mathbb{G}=\{\phi_p : p = 1,\dots N\}$. We further introduce the hole space, $\mathbb{H} = \mathbb{C}\cup\mathbb{A}$, and the particle space, $\mathbb{P} = \mathbb{A}\cup\mathbb{V}$.  The definitions of these spaces and the conventions for their indices are given in Table \ref{tab:orbital_space}.

\begin{table}
\centering
\caption{Definition of the orbital spaces employed in this work.}
\label{tab:orbital_space}
\begin{tabular}{l c c c}
\toprule
Space & Symbol & Indices & Definition \\
\midrule
Core & $\mathbb C$    & $m, n$ & occupied \\
 Active & $\mathbb A$ & $u, v, w, x, y, z$ & active \\
 Virtual & $\mathbb V$    & $e, f$ & unoccupied \\
 Hole & $\mathbb H$       & $i, j, k, l$ & $\mathbb H =\mathbb C \cup\mathbb A$ \\
 Particle & $\mathbb P$   & $a, b, c, d$ & $\mathbb P =\mathbb A \cup\mathbb V$ \\
 General & $\mathbb G$    & $p, q, r, s$ & $\mathbb G =\mathbb H \cup\mathbb V$ \\
\bottomrule
\end{tabular}
\end{table}

With the creation, $a^\dagger_p$, and annihilation, $a_q$, operators the excited configurations, $|\Phi_\mu\rangle$, in Eq.~(\ref{eq:ref}), can be expressed as
\begin{equation}
    |\Phi_\mu\rangle = \underbrace{a^\dagger_u \dots a^\dagger_z}_{N_A} \prod_{m\in\mathbb C} a^\dagger_m |0\rangle,
\end{equation}
where $|0\rangle$ denotes the physical vacuum. With the generalized normal ordering of an operator string, $\{a^\dagger_p a^\dagger_q \dots a_s a_r\}$, vacuum expectation values with respect to the generalized vacuum $|\Psi_0\rangle$ from Eq.~(\ref{eq:ref}), are always zero,
\begin{equation}
   \langle \Psi_0 | \{a^\dagger_p a^\dagger_q \dots a_s a_r\} | \Psi_0 \rangle = 0.
\label{eq:normal_ordering}
\end{equation}
Matrix elements evaluated with the generalized Wick theorem\cite{Kong2010_AlgebraicProof},
which contain contractions with active indices, yield expressions that depend on the one-particle, $\gamma^u_v$, or one-hole, $\eta^u_v$, density matrices:
\begin{align}
    \gamma^u_v &= \langle \Psi_0 | a^\dagger_u a_v | \Psi_0 \rangle, \\
    \eta^u_v &= \delta^u_v- \gamma^u_v.
\end{align}
In contrast to the normal ordering with a single-determinantal vacuum, multi-legged contractions between active indices appear. Those multi-legged contractions appear as contractions between even numbers, $2k$, of operators and require the evaluation of the $k$-body density cumulant. For example, a four-legged contraction yields the two-body cumulant
\begin{equation}
    \lambda^{uv}_{yz} =  \gamma^{uv}_{yz} - \gamma^u_y \gamma^v_z + \gamma^u_z \gamma^v_y,
\end{equation}
with the two-body RDM defined as
\begin{equation}
  \gamma^{uv}_{yz} = \langle \Psi_0 | a^\dagger_u a^\dagger_v a_z a_y | \Psi_0 \rangle.
\end{equation}
Expressions for higher-order cumulants are given, e.g., in Refs.~\citenum{Schaefer2020_GNO-RDMs} and \citenum{Kutzelnigg2010_spinfree}. 
Given the generalized Wick theorem and the condition from Eq.~(\ref{eq:normal_ordering}), we can express the Born--Oppenheimer electronic Hamiltonian, normal-ordered with respect to $|\Psi_0\rangle$ (denoted by the curly brackets), as
\begin{equation}
H = E_0 + \sum^\mathbb{G}_{pq} f^q_p \{a^\dagger_p a_q\} + \frac{1}{4} \sum^
\mathbb{G}_{pqrs} v^{rs}_{pq} \{a^\dagger_p a^\dagger_q a_s a_r\},
    \label{eq:NO_H}
\end{equation}
with the reference energy $E_0=\langle \Psi_0 | H | \Psi_0 \rangle$ and the generalized Fock matrix
\begin{equation}
     f^q_p = h^q_p + \sum^\mathbb{C}_n v^{qn}_{pn} + \sum^\mathbb{A}_{uv} v^{qv}_{pu} \gamma^u_v,
\label{eq:fock}     
\end{equation}
where $h^q_p=\langle \phi_q| h | \phi_p\rangle$ are the one-body matrix elements and $v^{rs}_{pq}=\langle pq || rs \rangle$ are the antisymmetrized two-body integrals. Following Evangelista and Li,\cite{Evangelista2015_MRDSRG,Evangelista2016_RelaxedDSRG,Evangelista2019_DSRG-Rev} we work in a semicanonical orbital basis where the core-core, active-active, and virtual-virtual blocks of the generalized Fock matrix, Eq.~(\ref{eq:fock}), are diagonal. We note that this ensures orbital invariance of the amplitude equations as shown in Ref.~\citenum{Evangelista2016_RelaxedDSRG}.

\subsection{Similarity renormalization group with nonunitary transformations}

We extend the IM-SRG method from a unitary similarity transformation to a nonunitary one. This will provide the basis for rationalizing the renormalization of CC amplitudes, as CC methods inherently rely on nonunitary transformations.
Our motivation for employing a nonunitary transformation is that in single-reference coupled cluster theory, the expansion of the Hamiltonian converges more quickly.\cite{Evangelista2011_AlternativeSinglereference}
However, it remains unclear whether this observation will in general hold for the multireference case.

For simplicity, we limit ourselves to the single-reference formulation here, but demonstrating the validity of a nonunitary SRG approach will enable us to adapt Li and Evangelista's unitary multireference DSRG formulation to a nonunitary MRCC method. 
The idea of the SRG is to generate a flow that drives the Hamiltonian towards a diagonal form.
We start with the continuous SRG transformation in terms of the flow parameter $s$ of the normal-ordered Hamiltonian
\begin{equation}
    H(s) = \Omega(s) H(0) \Omega^{-1}(s) = H_0(s) + H_1(s) + H_2(s) + H_3(s) + \dots\ ,
    \label{eq:srg_H}
\end{equation}
where on the right-hand side $H(s)$ is expanded in terms of $n$-body operators, $H_n(s)$.
We differentiate with respect to $s$ to obtain the flow equation for the Hamiltonian
\begin{equation}
\begin{aligned}
    \frac{\mathrm{d} H(s)}{\mathrm{d} s} = \frac{\mathrm{d} \Omega(s)}{\mathrm{d} s} H(0) \Omega^{-1}(s)  + \Omega(s)  H(0) \frac{\mathrm{d} \Omega^{-1}(s)}{\mathrm{d} s}.
\end{aligned}
\end{equation}
To obtain the final form of the flow equation, we exploit the identity 
\begin{equation}
    \frac{\mathrm{d} \Omega^{-1}(s)}{\mathrm{d} s} = -\Omega^{-1}(s)\frac{\mathrm{d} \Omega(s)}{\mathrm{d} s}\Omega^{-1}(s),
\end{equation}
in such a way that we obtain, after some manipulations,
\begin{equation}
    \frac{\mathrm{d} H(s)}{\mathrm{d} s} = [\zeta(s), H(s)],
    \label{eq:flow}
\end{equation}
where we introduced the generator of the flow as
\begin{equation}
    \zeta(s) = \Omega^{-1}(s)\frac{\mathrm{d} \Omega(s)}{\mathrm{d} s}.
\end{equation}
Since $\Omega(s)$ is a priori unknown, the key step is selecting a generator, $\zeta(s)$, that effectively decouples the off-diagonal elements of the Hamiltonian from the ground state.
In the IM-SRG, the goal is to decouple the occupied-virtual, $H^{ij\dots}_{ab\dots}(s)$, and virtual-occupied, $H^{ab\dots}_{ij\dots}(s)$, off-diagonal elements given antihermitian generator that generates the symmetric unitary transformation.
However, in CC theory, we recognize that it is sufficient to decouple the occupied-virtual blocks (or the virtual-occupied ones) of the Hamiltonian from the ground state, given an asymmetric-nonunitary similarity transformation.
To achieve the desired CC-like decoupling, we introduce the canonical \textit{asymmetric} Wegner\cite{Wegner1994_FlowequationsHamiltoniansa,Kehrein2006_FlowEquation} generator
\begin{equation}
    \zeta(s) = [H^\mathrm{d}(s), H^\mathrm{od}(s)],
    \label{eq:generator}
\end{equation}
where $H^\mathrm{d}(s)$ is the diagonal part of the Hamiltonian 
\begin{equation}
    H^\mathrm{d}(s) = \sum_p^\mathbb{G} \epsilon_p(s) \{a^\dagger_p a_p\},
\end{equation}
with $\epsilon_p(s) = f_{pp}(s)$, and the occupied-virtual off-diagonal part of the Hamiltonian is given by
\begin{equation}
    H^\mathrm{od}(s) = \sum_i^\mathbb{H} \sum_a^\mathbb{P} f^i_a(s) \{a^\dagger_a a_i\} + \frac{1}{4} \sum_{ij}^\mathbb{H} \sum_{ab}^\mathbb{P} v^{ij}_{ab}(s)\{a^\dagger_a a^\dagger_b a_j a_i\} + \dots\ .
    \label{eq:H_od}
\end{equation}
To obtain the asymmetric transformation, we do not include the hermitian conjugate in Eq.~(\ref{eq:H_od}). 
With Wick's theorem, we can evaluate the commutator from Eq.~(\ref{eq:generator}) according to
\begin{equation}
    \zeta(s) = -\sum_i^\mathbb{H} \sum_a^\mathbb{P} f^i_a(s) \Delta_a^i(s) \{a^\dagger_a a_i\} - \frac{1}{4} \sum_{ij}^\mathbb{H} \sum_{ab}^\mathbb{P} v^{ij}_{ab}(s) \Delta^{ij}_{ab}(s) \{a^\dagger_a a^\dagger_b a_j a_i\} - \dots\ ,
\end{equation}
where $\Delta^{ij\dots}_{ab\dots}(s)$ are the M{\o}ller--Plesset denominators
\begin{equation}
    \Delta^{ij\dots}_{ab\dots}(s) = \epsilon_i(s) + \epsilon_j(s) + \dots -\epsilon_a(s) - \epsilon_b (s) - \dots\ .
\end{equation}
To demonstrate the decoupling, we perform a perturbative analysis and follow the steps of Kehrein\cite{Kehrein2006_FlowEquation} and Evangelista\cite{Evangelista2014_DSRG1}. We partition the initial Hamiltonian, i.e., at $s=0$, as
\begin{equation}
    H(0,\lambda) = H_\mathrm{d}(0) + \lambda H^{(1)}(0),\quad H^{(1)}(0) = H(0) - H_\mathrm{d}(0),
\end{equation}
such that we can expand $H(s,\lambda)$ in a power series
\begin{equation}
    H(s,\lambda) =  H_\mathrm{d}(0) + \lambda H^{(1)}(s) + \lambda^2 H^{(2)}(s) + \dots\ .
\end{equation}
We can then split up the flow equations, Eq.~(\ref{eq:flow}), into the one-body and two-body contributions, see Eq.~(\ref{eq:srg_H}), and analyze them in powers of $\lambda$. The first-order contribution linear in $\lambda$ gives\cite{Evangelista2014_DSRG1,Hergert2016_InmediumSimilarity}
\begin{align}
    \frac{\mathrm{d} }{\mathrm{d} s } f_a^{i,(1)}(s) &= - \left(\Delta_a^i(0)\right)^2  f_a^{i,(1)}(s),\\
    \frac{\mathrm{d}}{\mathrm{d} s }  v_{ab}^{ij,(1)}(s) &= - \left(\Delta_{ab}^{ij}(0)\right)^2  v_{ab}^{ij,(1)}(s).
\end{align}
We can solve these differential equations to obtain the solutions as\cite{Kehrein2006_FlowEquation}
\begin{align}
    f^{i,(1)}_a(s) &= f^{i,(1)}_a(0)\mathrm{e}^{-s\left( \Delta^{i}_{a}(0) \right)^2},
    \label{eq:decoupling1}
    \\
    v^{ij,(1)}_{ab}(s) &= v^{ij,(1)}_{ab}(0)\mathrm{e}^{-s\left( \Delta^{ij}_{ab}(0) \right)^2}.
    \label{eq:decoupling2}
\end{align}
This shows that, in analogy to the unitary case,\cite{Kehrein2006_FlowEquation,Evangelista2014_DSRG1,Hergert2016_InmediumSimilarity} the generator in Eq.~(\ref{eq:generator}) leads to an exponentially decaying decoupling of the off-diagonal occupied-virtual elements of the Hamiltonian from the ground state.  
Specifically, the equations show that the transformation is a proper renormalization in the sense that for a finite $s$, the matrix elements are left nearly unchanged if $|\Delta^{i\dots}_{a\dots}|<s^{-\frac{1}{2}}$, but for large differences, the matrix elements are decoupled.

We note that if the occupied-virtual blocks of the transformed Hamiltonian, Eq.~(\ref{eq:srg_H}), are zero in the limit
\begin{equation}
    \lim_{s\rightarrow\infty} H_{ab\dots}^{ij\dots}(s) = 0,
\end{equation}
then $\lim_{s\rightarrow\infty}E_0(s)$ will correspond to the exact ground-state energy. 

To properly formulate a multireference version of IM-SRG based on nonunitary transformations, we would need to follow the steps described in Ref.~\citenum{Hergert2016_InmediumSimilarity}. However, since this is not the focus of this work, we proceed in the next section by adapting the multireference DSRG theory of Evangelista and Li\cite{Evangelista2015_MRDSRG,Evangelista2016_RelaxedDSRG} to the ic-MRCC ansatz. 
For completeness, we note that, as in the IM-SRG approach, we do not formulate a proper decomposition of the Hamiltonian in the sense that $H(s)=H_\mathrm{d}(s) + H_\mathrm{od}(s)$. Hence, the approach is an approximation to the SRG method. 

\subsection{Renormalized ic-MRCC}
To introduce the ric-MRCC method, we first write CC theory as an effective Hamiltonian\cite{Primas1963_EffectiveHamil} theory in the spirit of the SRG approach from the previous section. Note that compared to Eq.~(11), we adopt the sign convention that is typical for CC methods in the following, whereas we relied on the sign convention typical in the SRG literature in Eq.~(11.).
Hence, we again write the parametrization of the Hamiltonian as a continuous nonunitary similarity transformation
\begin{equation}
    H \longrightarrow \Bar{H}(s) = \Omega^{-1}(s) H \Omega(s),
    \label{eq:rCC_transformation}
\end{equation}
with $\Omega(0)=\identity$, and for $s\rightarrow\infty$ we recover the exact CC transformation. Our parametrization for the operator $\Omega(s)$ is the CC-like wave operator
\begin{equation}
    \Omega(s) = \mathrm{e}^{T(s)},
    \label{eq:wave_operator}
\end{equation}
and the amplitudes can be expanded in terms of $n$-body operators
\begin{equation}
    T(s) = \sum_n^{N_\mathrm{el}} T_n(s).
\end{equation} 
where $N_\mathrm{el}$ denotes the number of electrons in the system. The $n$-body cluster operator is defined as
\begin{equation}
    T_n(s) = \frac{1}{(n!)^2} \sum_{ij\dots}^\mathbb{H}  \sum_{ab\dots}^\mathbb{V} t^{ij\dots}_{ab\dots}(s) \{a^\dagger_a a^\dagger_b \dots a_j a_i\}.
    \label{eq:cluster_op}
\end{equation}
$t^{ij\dots}_{ab\dots}(s)$ are the renormalized CC amplitudes, and we explicitly set those amplitudes to zero that contain only active indices
\begin{equation}
    t^{xy\dots}_{uv\dots}(s)=0,
\end{equation}
because this simplifies the equations and their effects can be accounted for by relaxing the reference coefficients in Eq.~(\ref{eq:ref}), see, e.g., Ref.~\citenum{Evangelista2016_RelaxedDSRG}.
Hence, we can evaluate Eq.~(\ref{eq:rCC_transformation}) by applying the CC-similarity transformation to the normal-ordered Hamiltonian in Eq.~(\ref{eq:NO_H}) according to
\begin{equation}
    \Bar{H}(s) =  \mathrm{e}^{-T(s)} H(s) \mathrm{e}^{T(s)} = \Bar{H}_0(s) + \sum^\mathbb{G}_{pq} \Bar{H}^q_p \{a^\dagger_p a_q\} + \frac{1}{4} \sum^\mathbb{G}_{pqrs} \Bar{H}^{rs}_{pq} \{a^\dagger_p a^\dagger_q a_s a_r\}  + \dots.
    \label{eq:H_bar}
\end{equation}
Except for the parametric dependence on the parameter $s$, $\Omega(s)$ is exactly the wave operator of the standard ic-MRCC theory. The amplitudes of the wave operator can be obtained in two ways: either by solving the projected equations or by solving the many-body equations (note that in the single-reference formulation, those equations are equivalent). For a comparative analysis, see Refs.~\citenum{Datta2011_StatespecificPartially} and \citenum{lechner2023_icMRCC-Thesis}. In this work, we rely on the many-body formulation, also employed by Evangelista and Li in the multireference version of the DSRG\cite{Evangelista2015_MRDSRG}. In the many-body formulation, the conditions for decoupling the off-diagonal elements of the Hamiltonian from the ground state are given as
\begin{equation}
   \mathrm{lim}_{s\rightarrow\infty} \Bar{H}_{ab\dots}^{ij\dots}(s) = 0,
   \label{eq:mb_condition}
\end{equation}
which determine the amplitudes $t^{ij\dots}_{ab\dots}(s)$. 
Evangelista showed in his original work on the DSRG in Ref.~\citenum{Evangelista2014_DSRG1}, that amplitude-update equations can be derived by requiring that the elements $\Bar{H}_{ij\dots}^{ab\dots}(s)$ are driven by the so-called sources $r_{ab\dots}^{ij\dots}(s)$, given as\cite{Evangelista2014_DSRG1,Evangelista2016_RelaxedDSRG}
\begin{equation}
    \Bar{H}_{ab\dots}^{ij\dots}(s) \leftarrow r_{ab\dots}^{ij\dots}(s) = [ \Bar{H}_{ab\dots}^{ij\dots}(s) + t^{ij\dots}_{ab\dots}(s) \Delta^{ij\dots}_{ab\dots} ] \mathrm{e}^{-s(\Delta^{ij\dots}_{ab\dots} )^2},
    \label{eq:source_operator}
\end{equation}
with $\Delta^{ij\dots}_{ab\dots}=\Delta^{ij\dots}_{ab\dots}(0)$.
These equations determine the conditions for the CC amplitudes. To obtain an algorithm for updating the amplitudes, we note that in the limit $s\rightarrow\infty$, we can employ the standard CC identity
\begin{equation}
    ^\mathrm{new}t^{ij\dots}_{ab\dots}\Delta^{ij\dots}_{ab\dots} = {^\mathrm{old}\Bar{H}_{ab\dots}^{ij\dots}} + {^\mathrm{old}t^{ij\dots}_{ab\dots}} \Delta^{ij\dots}_{ab\dots}.
\end{equation}
This enables us to manipulate the left-hand side of Eq.~(\ref{eq:source_operator}) to obtain
\begin{equation}
   \lim_{s\rightarrow\infty} r_{ab\dots}^{ij\dots}(s) = \lim_{s\rightarrow\infty}\left(^\mathrm{old}\Bar{H}_{ab\dots}^{ij\dots}(s) + ^\mathrm{old}t^{ij\dots}_{ab\dots}(s) \Delta^{ij\dots}_{ab\dots} - ^\mathrm{new}t^{ij\dots}_{ab\dots}(s)\Delta^{ij\dots}_{ab\dots}\right)=0.
\end{equation}
Consequently, the conditions from Eq.~(\ref{eq:source_operator}) lead to the following amplitude-update equations\cite{Evangelista2014_DSRG1,Evangelista2016_RelaxedDSRG}
\begin{equation}
    ^\mathrm{new}t^{ij\dots}_{ab\dots}(s) = \left( ^\mathrm{old}\Bar{H}_{ab\dots}^{ij\dots}(s) +  ^\mathrm{old}t^{ij\dots}_{ab\dots}(s) \Delta^{ij\dots}_{ab\dots} \right) \frac{1 -  \mathrm{e}^{-s(\Delta^{ij\dots}_{ab\dots} )^2}}{\Delta^{ij\dots}_{ab\dots} },
\end{equation}
which reduce to the standard ic-MRCC equations in the limit $s\rightarrow\infty$. 
The only difference so far to the MR-DSRG formulation is that, instead of the wave operator $\Omega(s)$ in Eq.~(\ref{eq:wave_operator}), the DSRG relies on a unitary wave operator $\Omega(s) = \mathrm{e}^{T(s)-T^\dagger(s)}$.

We highlight here our reasons for introducing the renormalization-like condition: the many-body condition without renormalization, i.e., Eq.~(\ref{eq:mb_condition}), leads to poor convergence and even divergence when applied to the ic-MRCC theory\cite{Datta2011_StatespecificPartially}. This is due to numerical instabilities introduced by very small M{\o}ller--Plesset denominators. Introducing a renormalization factor, also called regularization factor\cite{Lee2018_RegularizedOrbitalOptimized,Shee2021_RegularizedSecondOrder,Battaglia2022_RegularizedCASPT2}, suppresses these numerical instabilities. In practice, we optimize the amplitudes at a finite value of $s$, usually $s=0.5$, which leads to a numerically stable method. The price to pay for the renormalization is that, for finite $s$, the method will not converge to the FCI result, even in the case of a full ric-MRCC ansatz with a maximum excitation rank equal to the number of particles. 

\subsection{Double commutator approximation}

We proceed now with the approximate evaluation of the many-body equations $\Bar{H}_{ab\dots}^{ij\dots}(s)$ in Eq.~(\ref{eq:mb_condition}). In particular, we assume here the singles and doubles approximation, ric-MRCCSD, where, as in conventional CCSD, the cluster operator, Eq.~(\ref{eq:cluster_op}), is truncated after the second order.
We rely on the \textsc{Wick\&d} program\cite{Evangelista2022_Wicked} for automatically generating all equations. The source code of our work, including the equation generation, is available on Zenodo\cite{ricmrcc}.

To start with, we expand the similarity transformed Hamiltonian from Eq.~(\ref{eq:H_bar}) with the BCH formula as
\begin{equation}
    \Bar{H}(s) =  \mathrm{e}^{-T(s)} H \mathrm{e}^{T(s)} = H + [H, T(s) ] + \frac{1}{2} [[H, T(s) ], T(s)] + \dots.
    \label{eq:bch}
\end{equation}
For a single-reference system, it is well established that the BCH expansion in Eq.(\ref{eq:bch}) terminates naturally after the fourth commutator. However, with the multi-determinantal reference of Eq.(\ref{eq:ref}), the expansion truncates much later\cite{Hanauer2011_PilotApplications}. Hence, similar to related methods such as DSRG, CT, IM-SRG, and unitary CC, one must either truncate the BCH expansion at a designated order or apply the recursive commutator expansion \cite{Yanai2006_CanonicalTransformation}. Evangelista, in the context of the DSRG, chose the latter approach.\cite{Evangelista2014_DSRG1} Yet, he and Gauss\cite{Evangelista2012_ApproximationSimilaritytransformed} highlighted that this approach could introduce considerable inaccuracies due to mismatched prefactors in the double commutator. Consequently, in this work, we will explore an alternative approximation strategy. 

The simplest approximation would be to keep only the linear commutator. However, Evangelista and Gauss have also shown\cite{Evangelista2011_OrbitalinvariantInternally} that this approach does not achieve high accuracy. 
Nonetheless, calculating the exact double commutator is impractical except for very small systems. 
Evaluating only the relevant occupied-virtual elements, the computational cost for evaluating the double commutator scales as $N^8$, where $N$ can stand for the number of occupied, $O$, virtual, $V$, or active, $A$, orbitals (cf.\ Table \ref{tab:orbital_space}). In addition, it involves a large prefactor due to the need to evaluate 10,146 terms.
For comparison, the linear commutator scales as $N^6$ and involves only 361 terms. In the framework of general normal ordering, the increase in complexity compared to SRCC methods can be attributed to the fact that the set of active orbitals, $\mathbb{A}$, is nonempty. 
Consequently, contractions that include amplitudes with active indices become the critical bottleneck in the computation of these commutators. For clarity, we note here that by \textit{amplitudes} we are always referring to the cluster amplitudes, $t^{p\dots}_{q\dots}$.

We chose to employ two different approximations to the double commutator: one for the contribution to the residuals and a different approximation to the contribution to the energy -- a strategy also previously employed by Black and K{\"o}hn\cite{Black2019_LinearQuadratic}. This choice is driven by practical considerations. Given that evaluating all terms is impractical, our goal is to evaluate as many terms as feasible while ensuring that the computational scaling remains reasonable.

If we assume that we have chosen an active space that is reasonably well separated from the external space, then the coupling between the active space and the external space is reasonably small.\cite{Faulstich2019_TCCAnalysis} Hence, we can assume that the amplitudes containing active and external indices are also likely to be small.
It is, of course, not always possible to find an active space that fulfills this requirement, but nonetheless, it will be the starting point for our approximations: 
Our key approximation to the double commutator is based on neglecting all contractions involving two amplitudes with active indices. Hence, we retain the contractions between two amplitudes without active indices (i.e., the standard CC contractions) and between amplitudes without and with active indices. This reduces the number of terms required to evaluate the double commutator to 513.
We note that this might make the method more sensitive to the choice of the active space.
However, our AutoCAS algorithm based on quantum information metrics can help to select appropriate active spaces reliably.\cite{Vera2016_delicate,Stein2016_AutomatedSelection,Stein2017_Chimia,Stein2019_AutoCAS-Implementation}

The second approximation is based on the observation by Hanauer and K{\"o}hn \cite{Hanauer2011_PilotApplications,Hanauer2012_PerturbativeTreatment} that amplitudes with several active indices are more likely to be redundant. 
In particular, they observed that for a limited set of examples, doubles amplitudes with three active indices and triples amplitudes with more than three indices may be redundant.\cite{Hanauer2011_PilotApplications,Hanauer2012_PerturbativeTreatment}
Therefore, we also neglect the contractions involving the two-body cumulant to the double commutator since these involve contractions with amplitudes with three active indices, which reduces the number of terms further to 463. Our approximations are summarized in Table \ref{tab:approx}, and we note that the approximate equations scale according to $N^6$.

\begin{table}[]
    \centering
    \begin{tabular}{l c}
    \toprule
        Contribution & Approximations \\
    \midrule 
      $R$: $[[H,T_\mathrm{SD}],T_\mathrm{SD}]]_\mathrm{approx}$ & $t^{p\dots}_{\textcolor{SchoolColor}{\bm u}\dots}t^{\textcolor{SchoolColor}{\bm u}\dots}_{q\dots}=0$, $\lambda_{\textcolor{SchoolColor}{\bm u\bm v}}^{\textcolor{SchoolColor}{\bm w}x} t^{\textcolor{SchoolColor}{\bm u\bm v}}_{\textcolor{SchoolColor}{\bm w}p\dots}= 0$ \\
      $E$: $[[H,T_\mathrm{SD}],T_\mathrm{SD}]]_\mathrm{approx}$ & $\lambda_4=0$,  $t_{\textcolor{SchoolColor}{\bm u\bm v}}^{\textcolor{SchoolColor}{\bm w}p} t^{\textcolor{SchoolColor}{\bm u\bm v}}_{\textcolor{SchoolColor}{\bm w}q}= 0$\\
      $E$: $[H,T_3]_\mathrm{approx}$ & $\lambda_4=0$\\
      $E$: $[[H,T_\mathrm{SD}],T_3]]_\mathrm{approx}$ & $\lambda_4=0$,  $t_{\textcolor{SchoolColor}{\bm u\bm v}p}^{\textcolor{SchoolColor}{\bm w\bm x}q} t^{\textcolor{SchoolColor}{\bm u\bm v}p}_{\textcolor{SchoolColor}{\bm w\bm x}q}= 0$\\
    \bottomrule
    \end{tabular}
    \caption{Computational scaling of the linear and double commutator contributions to the residuals, $R$, and the energy, $E$. The singles and doubles amplitudes (SD) are defined as $T_\mathrm{SD}=T_1+T_2$, and active indices are highlighted in bold-faced blue letters.}
    \label{tab:approx}
\end{table}

We further analyze the contributions of the linear and double commutators to the energy $\Bar{H}_0(s)$, which is the zero-body contribution in Eq.~(\ref{eq:H_bar}). Compared to the single-reference case, the energy evaluation may introduce a bottleneck. The scalar contribution from the single commutator already requires the evaluation of the three-body cumulant, $\lambda_3$, leading to a $N^7$ scaling. Evangelista showed that for the MR-DSRG method, its contribution is significant\cite{Evangelista2015_MRDSRG}. Evaluating the scalar contribution to the double commutator requires even the four-body cumulant, $\lambda_4$. Since evaluating $\lambda_4$ would severely limit the size of the active space, we will neglect its contribution entirely.
The linear commutator is evaluated exactly and contributes 24 terms to the energy, where two terms require the three-body cumulant and scale as $N^7$. The contributions of the double commutator involve 182 terms, with three terms that include the four-body cumulant.
To further minimize computational demand in evaluating the double commutator, we exclude contractions involving two amplitudes, each with three active indices. This reduces the number of terms to 100. 

\begin{table}[]
    \centering
    \begin{tabular}{l c c}
    \toprule
        Contribution & Scaling & Num.\ terms\\
    \midrule
        $R$: $[H,T_\mathrm{SD}]$ & $A^2V^4$, $C^2V^4$ & 361\\
        $E$: $[H,T_\mathrm{SD}]$ & $A^4V^2$, $A^6V$ & 24\\
        $R$: $[[H,T_\mathrm{SD}],T_\mathrm{SD}]]_\mathrm{approx}$ & $A^2V^4$, $C^2V^4$, $CAV^4$ & 463\\
        $E$: $[[H,T_\mathrm{SD}],T_\mathrm{SD}]]_\mathrm{approx}$ & $A^4V^2$, $CA^3V^2$, $A^6V$ & 100\\
        $T_3$: $[H,T_\mathrm{SD}]$ & $A^5V^2$, $A^4V^3$, $CA^3V^3$ & 100\\
        $E$: $[H,T_3]$ & $A^5V^3$ & 23 \\
        $E$: $[[H,T_\mathrm{SD}],T_3]]_\mathrm{approx}$ & $C^2A^3V^2$ & 16 \\
    \midrule
        $R$: $[[H,T_\mathrm{SD}],T_\mathrm{SD}]]$ & $A^4V^4$ & 10,146\\
    \bottomrule
    \end{tabular}
    \caption{Number of terms and computational scaling of the linear and double commutator contributions to the residuals, $R$, and the energy $E$. The singles and doubles amplitudes (SD) are defined as $T_\mathrm{SD}=T_1+T_2$. $A$ refers to the number of active orbitals, $C$ is the number of core orbitals, and $V$ is the number of virtual orbitals.}
    \label{tab:scaling}
\end{table}

We provide details of the scaling for the contributions to the residuals, $R$, and the energies, $E$, in Table \ref{tab:scaling}, where we relied on the factorization of the NumPy einsum function with the `optimal' setting\cite{harris2020array}. 
We focus on the highest power of $V$ as it typically exceeds $C$ and $A$ in size. 
Additionally, the number of core orbitals is typically larger than the number of active orbitals.
Given these conditions, the predominant scaling factor for the residuals of the linear commutator is $C^2V^4$. For energy contributions, the scaling of the single commutator, due to the three-body cumulant, is $A^6V$. 
However, given that $A$ is significantly smaller than $V$, the most expensive equation scales as $A^4V^2$.
The contribution to the residuals of the double commutator is $C^2V^4$ since, here, we neglect the three-body cumulant.
Consequently, except for the contribution of the three-body cumulant to the energy, we retain the desirable scaling of $C^2V^4$, as in standard CCSD, with an even more favorable scaling than MR-LDSRG(2), since for the recursive commutator expansion, the entire similarity transformed Hamiltonian must be evaluated.

\subsection{Perturbative triples correction}

In this section, we introduce the perturbative triples approximation derived by Hanauer and K{\"o}hn in Ref.~\citenum{Hanauer2012_PerturbativeTreatment}. However, we rely on the many-body instead of the projected approach.

We evaluate the triples amplitudes with the contribution of the linear commutator to the third order, i.e.,
\begin{equation}
    t^{ijk}_{abc} = \Bar{H}_{abc}^{ijk}(s) \frac{1 -  \mathrm{e}^{-s(\Delta^{ijk}_{abc} )^2}}{\Delta^{ijk}_{abc} },
\end{equation}
where $\Bar{H}_{abc}^{ijk}(s)$ is evaluated with the converged singles and doubles amplitudes according to
\begin{equation}
    \Bar{H}_{abc}^{ijk}(s) \approx  [H, T_2(s)]^{ijk}_{abc}.
\end{equation}
This approach is related to the single-reference CCSD[T] method.\cite{Urban1985_TowardsFullCCSDT}
To evaluate the contribution of the triples amplitudes to the energy, we partition the Hamiltonian as in DSRG-MRPT
\begin{equation}
    H = H^{(0)} + H^{(1)},
\end{equation}
with the zeroth-order Hamiltonian consisting of the diagonal part,
\begin{equation}
    H^{(0)} = E_0 + \sum_p^\mathbb{G} \epsilon_p \{a^\dagger_p a_p\},
\end{equation}
and the perturbation Hamiltonian consisting of all off-diagonal contributions
\begin{equation}
    H^{(1)} = H - H^{(0)}.
\end{equation}
Alternatively, one could explore different partitionings of the Hamiltonian.\cite{Hanauer2012_PerturbativeTreatment}
In this work, we focus on the direct contribution of the triples amplitudes to the energy, a term that is not present in single-reference CC. We refer to this approach as ric-MRCCSD[T]. The direct contribution to the energy is written as the scalar component of the following expression\cite{Hanauer2012_PerturbativeTreatment}
\begin{equation}
    \Delta E_\mathrm{[T]} =  [H^{(1)}, T_3] + [[H^{(1)}, T_1], T_3] + \frac{1}{2}[[H^{(1)}, T_2], T_3] +  \frac{1}{2}[[H^{(1)}, T_3], T_2]  \Big{|}_0,
    \label{eq:triples}
\end{equation}
where we omitted the dependence on $s$ for brevity.
To develop a computationally practical version of ric-MRCCSD[T], we have simplified the single and double commutators by setting the four-body cumulant to zero. The linear commutator is not approximated any further. However, in the evaluation of the double commutator, we excluded all contributions from contractions involving amplitudes with more than three active indices, in line with the findings of Hanauer and K{\"o}hn mentioned above.\cite{Hanauer2012_PerturbativeTreatment}
As a result, the number of terms for the linear commutator is 23 (omitting two terms that involve $\lambda_4$), and for the double commutator, we have reduced the number of terms from 637 to only 16.

We chose not to include the (T) correction derived by Hanauer and K{\"o}hn in our approach for the following reasons: if we categorize classes of amplitudes by their distinct combinations of active, core, and virtual indices, then the direct contribution to the energy will require only the triples amplitudes with at most two virtual indices and seven classes of amplitudes in total. 
By contrast, the (T) correction requires evaluating amplitudes with up to three virtual indices and of 15 different classes. Hence, the computational timing for evaluating these additional amplitudes is quite substantial.
Furthermore, the (T) contribution requires contractions involving three amplitudes -- derived from the Lambda equations -- which significantly increases the computational effort even further. We have performed some test calculations on the additional (T) energy contribution and found that it is an order of magnitude smaller than the [T] contribution. 
Therefore, we have chosen to omit the (T) terms and defer a more detailed investigation to future work.

To conclude this section, we analyze the scaling of the triples correction, provided in Table \ref{tab:scaling}: for the evaluation of the amplitudes, the computational scaling is $C^2A^3V^2$, that is, it has the same $N^7$ scaling as in the single-reference case. 
However, the evaluation of the energy scales as $A^5V^3$. Yet, the bottleneck involves a high power of active indices, and we assume that $A$ is reasonably small, resulting in the $C^2A^3V^2$ scaling.

\section{Computational Details}
\label{sec:comp_details_mrcc}

Unless stated otherwise, we relied on PySCF\cite{Sun2018_PySCF1,Sun2020_PySCF2} for Hartree--Fock, CASSCF, and NEVPT2 calculations. The TCC with singles and doubles (TCCSD), MR-LDSRG(2), ric-MRCCSD, and ric-MRCCSD[T] methods are implemented in the spin-orbital basis in our Python packages available via Zenodo\cite{casicc,ricmrcc}, where all equations were generated with \textsc{Wick\&d}.\cite{Evangelista2022_Wicked} The MR calculations were always converged below 10$^{-6}$~Ha, and, if not stated otherwise, all calculations employ a renormalization parameter $s=0.5$, given in Hartree atomic units. 
DMRG calculations were conducted with QCMaquis\cite{QCMaquis_2015} with a bond dimension, $m=2000$, and we employed the Fiedler ordering\cite{Legeza2003_OrderingOptimization,Reiher2011_Fiedler} for convergence acceleration. Internally-contracted MRCISD\cite{Werner1988_icMRCI,Shamasundar2011_NewIcMRCI} calculations were carried out with Molpro\cite{Werner2012_MolproGeneralpurpose} 2015 with the full valence active space including the Davidson correction\cite{Davidson1974_ConfigurationInteraction,Langhoff1974_ConfigurationInteraction} (ic-MRCISD+Q).

\begin{figure}
\centering
\includegraphics[width=0.29\textwidth]{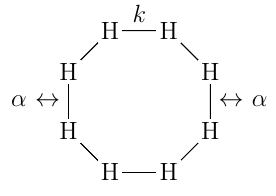}
\caption{Structure of the H$_8$ system. The $k$ parameter specifies the bond lengths of the H$_2$ units, which are fixed at 1~{\AA}. The $\alpha$ parameter denotes the displacement of the left and right H$_2$ units, varying between $-0.22$ and 0.22~{\AA}, with $\alpha=0$ representing the octagonal configuration.}
\label{fig:h8_system}
\end{figure}

The configuration of the H$_8$ system was chosen as in depicted in Fig.~\ref{fig:h8_system}, and we employed the 6-31G basis set\cite{Pople1980_self} with a CAS(8,8). We chose a CAS(4,4) with two orbitals each transforming according to the {$a_1$} and {$b_2$} irreducible representations of the C$_{2v}$ point group for the H$_2$O molecule, and we investigate the symmetric double dissociation with a bond angle of $104.5^\circ$.
We selected the cc-pVDZ basis set\cite{Dunning1989_gaussian} for the H$_2$O, F$_2$, and N$_2$ molecules, with a CAS(2,2) for F$_2$, and a full-valence CAS(10,8) for N$_2$. We selected this active space for N$_2$ since in Ref.~\citenum{Angeli2001_NelectronValence}, it was shown that this CAS provides the most accurate spectroscopic constants.

Calculations on the chromium dimer employed scalar relativistic effects through the X2C Hamiltonian\cite{Kutzelnigg2005_QuasirelativisticTheory,Kutzelnigg2006_QuasirelativisticTheory,Liu2007_QuasirelativisticTheory,Ilias2007_InfiniteorderTwocomponent}, the cc-pVDZ-DK basis set,\cite{Balabanov2005_SystematicallyConvergent} and the minimal CAS(12,12) consisting of the 4$s$ and 3$d$ orbitals selected according to symmetry labels, as detailed in the Python script available on Zenodo\cite{ricmrcc}.
Errors of a method (`$\mathrm{M}$') at a given point $x$ on the potential energy surface (PES) were evaluated against a reference method (`$\mathrm{ref}$', DMRG or FCI) according to $\Delta E(x) = E_\mathrm{M}(x) - E_\mathrm{ref}(x)$. 
The non-parallelity error\cite{Li1995_NPE} (NPE) is defined as the difference between the largest and smallest deviation to the reference over the entire PES. 
Additionally, the mean absolute error (MAE) for a sample of $n$ data points is evaluated as $\mathrm{MAE} = \frac{1}{n} \sum_i^n |\Delta E_i|$. 

\section{Results}
\label{sec:results_mrcc}

\subsection{Renormalization and double commutator approximation}

\begin{figure}
    \centering
    \includegraphics[width=0.7\textwidth]{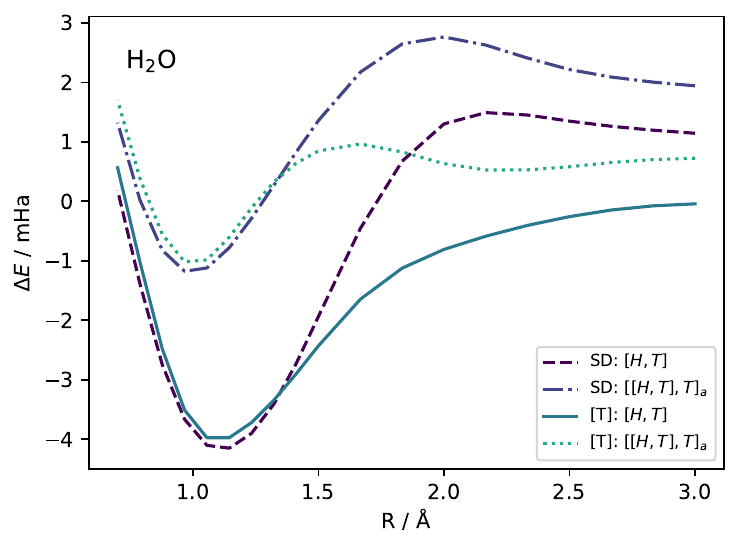}
    \caption{Energy deviation, $\Delta E$, of the H$_2$O dissociation in milliHartree (mHa) of the ric-MRCCSD (SD), and ric-MRCCSD[T] ([T]) with the linear ($[H,T]$) and approximated double ($[[H, T], T]_a$) commutator with a CAS(4,4) for the cc-pVDZ basis set, compared to DMRG($m=2000$).}
    \label{fig:comm_approx}
\end{figure}

First, we analyze whether approximating the double commutator can improve results compared to evaluating only the linear commutator. 
For that, we provide the error of the ric-MRCCSD and ric-MRCCSD[T] methods against DMRG for the double-bond dissociation of the water molecule in Fig.~\ref{fig:comm_approx}. 
Evaluating the approximate double commutator, denoted by $[[H, T], T]_a$, significantly reduced the error for both ric-MRCCSD and the ric-MRCCSD[T], which are abbreviated as SD and [T] in the figure, respectively.
Specifically, for ric-MRCCSD, the NPE is decreased by 1~milliHartree from 5.643~milliHartree to 4.541~milliHartree.
Including the [T]-correction, the error was reduced from 3.940~milliHartree to 2.716~milliHartree.
Fig.~\ref{fig:comm_approx} also shows that the largest changes in the error compared to the DMRG energy appear close to the equilibrium structure around 1~\AA, while the error stays constant in the dissociation limit.
Potentially, the reason is that H$_2$O is predominantly single-reference around the equilibrium geometry, which means that the active space most likely contains redundancies, i.e., orbitals with a natural occupation number that is close to zero.
Consequently, the renormalization most likely cancels out many states, and therefore, the approach suffers from a loss of accuracy.
These findings also align with the observations by Datta, Kong, and Nooijen.,\cite{Datta2011_StatespecificPartially} who found that they had to eliminate many amplitudes to converge the partially internally-contracted MRCCSD calculation close to the equilibrium.
We remark, however, that the ric-MRCC method converged smoothly in about 10 iterations or less along the entire PES.
Hence, we can confirm that the renormalization eliminates any convergence difficulties.

\begin{figure}
    \centering
    \includegraphics[width=0.7\textwidth]{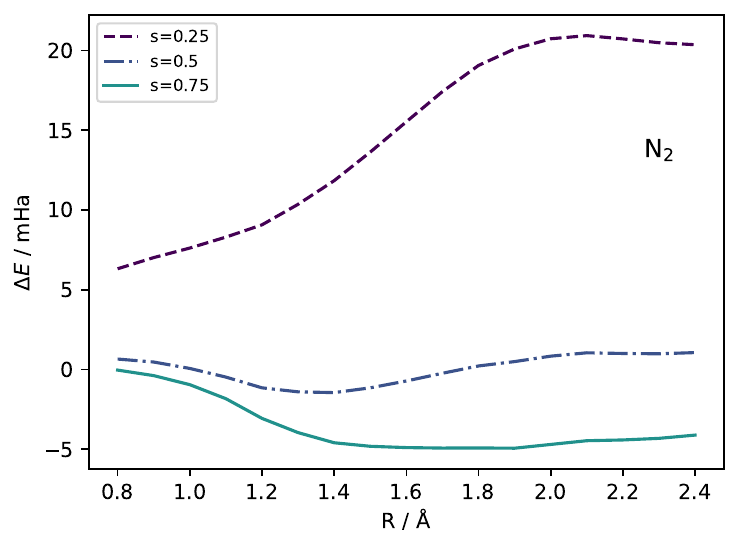}
    \caption{Energy deviation, $\Delta E$, of the N$_2$ dissociation in milliHartree (mHa) of the ric-MRCCSD[T] method compared to DMRG($m=2000$) with varying values for the renormalization parameter $s$ and a CAS(10,8) for the cc-pVDZ basis set.}
    \label{fig:s}
\end{figure}

However, we note that in the limit of $s\rightarrow\infty$, convergence may be reached in the dissociation limit but not around the equilibrium structure, aligning with findings from Ref.\citenum{Datta2011_StatespecificPartially}.
Furthermore, as extensively discussed by Evangelista and coworkers,\cite{Evangelista2014_DSRG1,Evangelista2016_RelaxedDSRG,Evangelista2019_DSRG-Rev} the choice of the renormalization parameter $s$ can significantly affect the quality of the results.
An $s$ value of 0.5 appears to offer an optimal balance.
We refer to Evangelista's work for an in-depth analysis, but we demonstrate for completeness that the same principle holds for ric-MRCC.
We compare the error of the PES of N$_2$ with ric-MRCCSD[T] for $s$ values of 0.25, 0.5, and 0.75, as shown in Fig.\ \ref{fig:s}.
The observations confirm that $s=0.5$ yields the most reliable results, with an NPE of 2.509 milliHartree, compared to 4.896~milliHartree for $s=0.75$ and 14.613~milliHartree for $s=0.25$.
We use $s=0.5$ for all subsequent calculations based on these findings.

\subsection{Potential energy curves of small molecules}

\begin{figure}
    \centering
    \includegraphics[width=\textwidth]{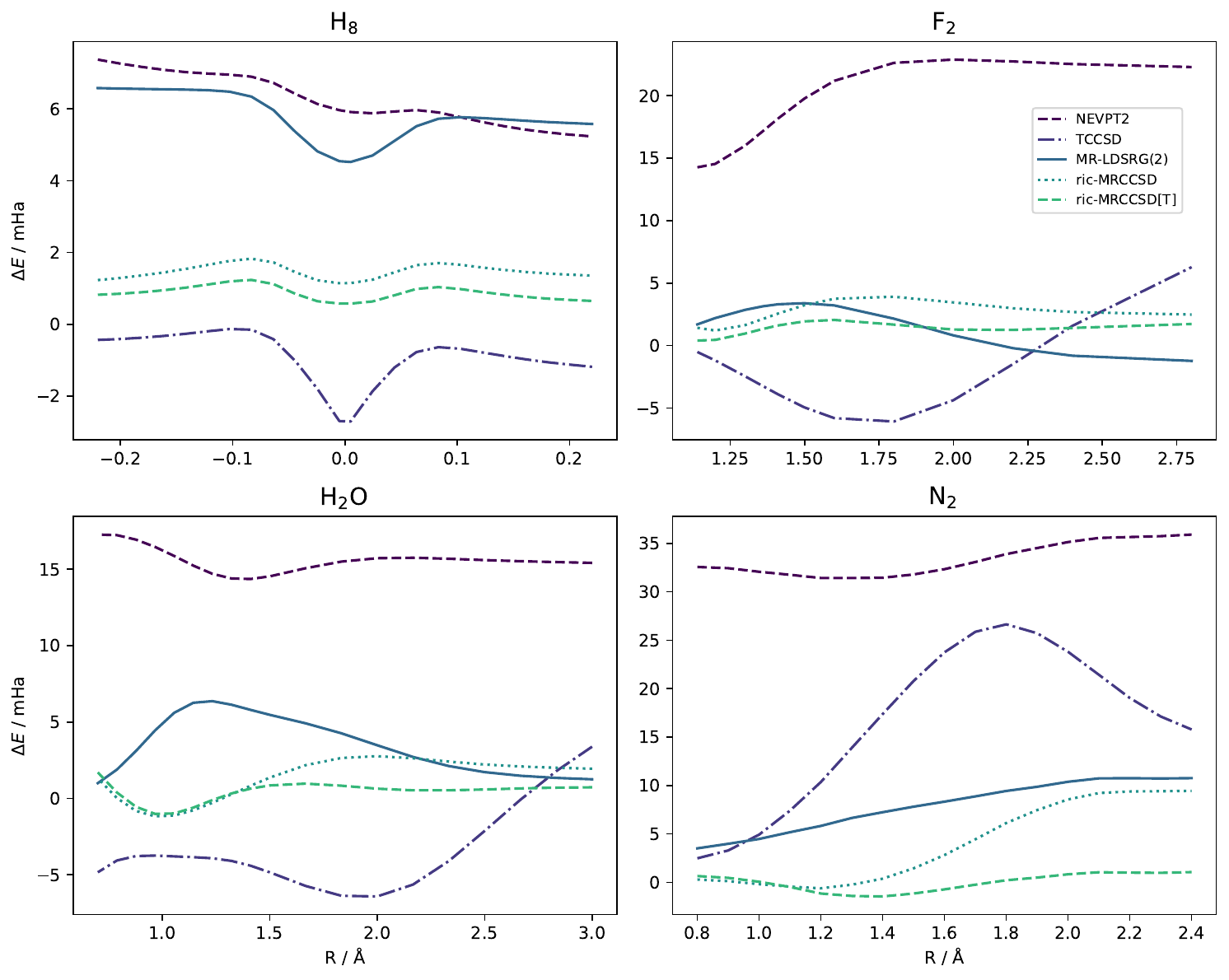}
    \caption{ 
   Energy deviations, $\Delta E$, in milliHartree (mHa) with respect to FCI (for H$_8$) and DMRG($m=2000$) for the dissociation of several small molecules. Shown are the results for NEVPT2, TCCSD, ric-MRCCSD, ric-MRCSSD[T], and MR-LDSRG(2) calculations. We employed the small 6-31G basis set for H$_8$ and the cc-pVDZ basis set for all other systems for the technical comparison.}
    \label{fig:combined}
\end{figure}

We proceed by comparing the NPEs of ric-MRCCSD and ric-MRCCSD[T] to that of TCCSD, MR-LDSRG(2), and NEVPT2 for H$_8$, F$_2$, H$_2$O, and N$_2$ with FCI-type results as a reference. 
The errors of the PESs are depicted in Fig.~\ref{fig:combined}, and the NPEs are listed in Table \ref{tab:small_molecules} where we also provide the MAE and maximum error (MAX) for every method.
We present only relative errors and not the PESs since the ric-MRCCSD, ric-MRCCSD[T], MR-LDSRG(2), and DMRG PESs are virtually indistinguishable.

First, we analyze Fig.~\ref{fig:combined}. Across all four graphs, NEVPT2 displays a large offset along the $y$-axis, and all NEVPT2 curves show some deviations from a constant horizontal line. 
TCCSD shows large deviations with notable local minima and maxima, sometimes reaching about 20 milliHartree energy deviation, 
which indicates an inaccurate description of the dissociation process. 
MR-LDSRG2 yields curves that display amplitudes that are always less than 10 milliHartree in magnitude, but there is a significant offset in the curve for H$_8$.
The ric-MRCC methods deliver the most consistent accuracy. 
In particular, the ric-MRCCSD[T] curves are nearly constant with small offsets along the $y$-axis. There are small, but recognizable deviations from a constant line.
Notably, for the H$_8$ system, the deviation of both ric-MRCC approaches is symmetric around the point of complete degeneracy ($R=0$~{\AA}), which is not the case for all other approaches where significant asymmetry can be observed.
This indicates that the ric-MRCC methods provide the most balanced description of correlation (for $R<-0.1$~{\AA}, the system is predominantly weakly correlated, while for $R>0.1$~{\AA}, 
strong correlation is also present to some degree). 

\begin{table}
    \centering
\begin{tabular}{l r r r r r}
\toprule
         Molecule & NEVPT2      &   TCCSD   &       MR-LDSRG(2) &   ric-MRCCSD      & ric-MRCCSD[T]       \\
\midrule          
Timing &&&&\\
\midrule           
         H$_8$   &    0.710   &          0.138   &          13.719   &         10.343   &         6.513   \\
         H$_2$O  &    0.030   &          0.846   &          11.902   &          4.144   &         1.688   \\
         F$_2$   &    0.020   &          3.001   &          26.380   &          7.186   &         0.294   \\
         N$_2$   &    2.552   &          1.902   &          48.331   &         16.497   &        31.988   \\
\midrule         
NPE  &&&&\\
\midrule         
         H$_8$   &    2.140   &          2.575   &          2.059   &          0.683   &          0.660\\
         H$_2$O  &    2.909   &          9.805   &          5.368   &          3.940   &          2.716\\
         F$_2$   &    8.620   &         12.350   &          4.616   &          2.707   &          1.648\\ 
         N$_2$   &    4.485   &         24.147   &          7.244   &         10.056   &          2.510\\ 
\midrule           
MAE 	 &	 4.538 &	 12.219 &	 4.821 &	 4.346 &	 1.883\\ 
MAX 	 &	 8.620 &	 24.147 &	 7.244 &	 10.056 &	 2.716 \\ 
\bottomrule
    \end{tabular}
    \caption{Timings per single point in seconds and non-parallelity errors in milliHartree for the potential energy curves studied in this work. We compare NEVPT2, ric-MRCCSD, ric-MRCCSD[T], MR-LDSRG(2), and TCCSD energies against FCI for H$_8$ in the 6-31G basis set and DMRG($m=2000$) with the cc-pVDZ basis set for the other systems. We also provide the mean absolute error (MAE) and maximum error (MAX) in milliHartree for the NPEs of each method.}
    \label{tab:small_molecules}
\end{table}

To analyze the results presented in Table \ref{tab:small_molecules}, we recall that Lyakh et al.\ \cite{Lyakh2012_MultireferenceNature} have established a guideline for assessing NEPs: NPEs between $1-5$~milliHartree are considered acceptable, those within the range of $5-10$~milliHartree are conditionally acceptable, and any error exceeding 10~milliHartree is deemed unacceptable.
Table \ref{tab:small_molecules} shows that NEVPT2 yields acceptable errors for three of the four molecules, with a MAX error of 8.620~milliHartree and an MAE of 4.538~milliHartree. TCCSD performs substantially worse, with unacceptable errors for N$_2$ and F$_2$. ric-MRCCSD and MR-LDSRG(2) perform similarly, with MAEs of around $4-5$~milliHartree, while MR-LDSRG(2) is slightly more accurate.

Compared to some results provided for the standard ic-MRCCSD method with the exact double commutator and without renormalization\cite{Evangelista2011_OrbitalinvariantInternally,Hanauer2011_PilotApplications} the approximate ric-MRCCSD performs significantly worse.
For instance, Hanauer and K{\"o}hn, and Evangelista and Gauss reported NPEs of less than 1~milliHartree for the dissociation of N$_2$ and H$_2$O.
However, the computational costs of the exact approaches and those presented in this work are very different.
To estimate the cost of the full double commutator, we performed a single-point calculation of the H$_4$ system\cite{Morchen2022_TCC,feldmann2024complete} in a CAS(4,4) and 6-31G basis set.
The calculation takes about 3006~s, while, by contrast, the calculation with the approximate double commutator takes only 6~s.

By including perturbative triples, we achieve excellent accuracy with an MAE of 1.883~milliHartree and a maximum error of 2.716~milliHartree for H$_2$O.
Evaluating the ic-MRCISD+Q PES for N$_2$ further demonstrates the reliability of our approach: we obtained an NPE of 1.767 milliHartree with the latter, and ric-MRCCSD[T] yields an NPE of 2.510 milliHartree.
Hence, with the [T] correction, the new approach nearly matches the accuracy of ic-MRCISD+Q but requires only the three-body RDM instead of the four-body RDM.

Moreover, Black and K{\"o}hn\cite{Black2019_LinearQuadratic} reported NPEs for the projected-residual ic-MRCCSD method applied to the N$_2$ dissociation with the same basis set and active space as in our calculations, but with frozen core electrons.
They observed NPEs of 6.810~milliHartree with only the linear commutator and 1.356~milliHartree when the double commutator is included.
Hence, the projected-residual ic-MRCCSD method provides significantly better results than ric-MRCCSD with an NPE of 10.056~milliHartree. 
However, in the projected-residual formulation, the redundant excitations must be numerically eliminated, which results in discontinuities in the PES when the full double commutator is not evaluated.\cite{Black2019_LinearQuadratic}
In contrast, we do not observe any discontinuities in the renormalized many-body residual approach.
Additionally, with the [T] correction, we can offset the errors from the approximation and renormalization, achieving results almost as accurate as those obtained with the exact double commutator but at a significantly lower computational cost. However, we cannot provide any theoretical reason for this error cancellation, and hence, it may be fortuitous.

To demonstrate the reasonable computational cost, we report the timings of a single-point calculation performed on a single core of an AMD(R) Epyc(TM) 75F3 central processing unit for all systems in Table \ref{tab:small_molecules}.
We highlight that NEVPT2 calculations were performed using highly optimized routines in PySCF, while all other calculations utilized our pilot implementation that is not optimized and carried out spin-orbital basis, resulting in an inefficient algorithm. 
Therefore, our timings are not competitive but serve as a rough estimate.
A spin-adapted implementation of MR-LDSRG(2) and ric-MRCC would accelerate calculations significantly.
Additionally, our current reliance on Numpy for factorizing equations without intermediates indicates considerable optimization potential.
We also note that for a fair comparison, we implemented an adaptive recursive commutator expansion for MR-LDSRG(2), where the number of nested commutators evaluated in every iteration depends on the current error of the optimization.
This adaptive algorithm significantly reduces the computational timings since, in the first iterations, three to four commutator evaluations are sufficient, while close to convergence, up to ten nested commutators must be evaluated.

When analyzing the timings in Table \ref{tab:small_molecules}, we observe, as expected, that NEVPT2 is the fastest method, except for H$_8$, where TCCSD is faster.
This is due to the small 6-31G basis set, which only gives 16 orbitals in total.
NEVPT2 is exceptionally fast when the active space is very small, as for H$_2$O, CAS(4,4) and F$_2$, CAS(2,2).
When the CAS becomes larger, the computational timings rapidly increase for NEVPT2 due to the need to evaluate contractions involving the four-body RDM.
The timings of the TCCSD method are all roughly in the same order of magnitude since the method does not directly depend on the size of the active space.

By contrast, the genuine MRCC methods are computationally more demanding. By comparing F$_2$, CAS (2,2), and N$ _ 2$, CAS (10,8), we can clearly see that the cost of the MR methods increases substantially with the size of the active space. Notably, the timing of NEVPT2 increased by about two orders of magnitude in this case, while the MR-LDSRG(2) and ric-MRCCSD timings roughly doubled. 
Similar to NEVTP2, the [T] correction cost is about two orders of magnitude more expensive. This can be ascribed to the contractions scaling as $A^5V^2$. 
However, for larger molecules, the contractions scale as $CA^3V^3$ because they will exceed the cost of those terms scaling as $A^5V^2$.

In addition, we also provide timings for a single-point calculation of the N$_2$ molecule for TCCSD, MR-LDSRG(2), and ric-MRCCSD with the CAS(10,8) in the cc-pVQZ basis. In this case, the number of virtual orbitals is significantly larger than that of core and active orbitals.
We performed the single-point calculations on 16 cores of an Intel(R) Xeon(R) Gold 6136 central processing unit. 
The TCCSD calculation took 1085~s, MR-LDSRG(2) 22899~s, ric-MRCCSD 1835.532~s, and the timing for the [T] correction was 1051~s. 
In this large basis set, MR-LDSRG(2) performed significantly worse compared to the other methods, which is due to the recursive approximation, which also requires the calculation of the entire Hamiltonian in the approximation of the nested commutator:
matrix elements with four virtual indices must be computed, which causes this poor efficiency.
Comparing ric-MRCCSD and TCCSD, timings for the large basis set differ by about a factor of two, compared to the factor of almost 100 for H$_8$ in the 6-31G basis set, see Table~\ref{tab:small_molecules}. This is because the scaling in the number of virtual orbitals dominates over the scaling in the number of active orbitals in the large basis set.

Therefore, we conclude that ric-MRCC outperforms MR-LDSRG(2) regarding computational speed while maintaining comparable accuracy regarding the NPEs for the systems studied here.
With the addition of the perturbative triples correction, we have enhanced the accuracy, bringing it nearly in line with ic-MRCISD+Q.
Moreover, already in its pilot implementation phase, ric-MRCC is promising in terms of computational cost with substantial potential for further performance improvements.

\subsection{Comparison to single-reference coupled cluster methods}

We now compare the MR methods to CCSD and CCSD(T) for equilibrium properties since the aim is to develop a method that addresses the deficiencies of SRCC in strongly correlated systems while achieving comparable accuracy in systems dominated by a single determinant. 

\begin{table}
\centering
\resizebox{\textwidth}{!}{%
\begin{tabular}{p{2.cm} R{1.1cm} R{2.cm} R{2.cm} R{2.8cm} R{3.2cm} R{3.0cm}}
\toprule
Molecule & CAS        & CCSD & CCSD(T)        & ric$-$MRCCSD & ric$-$MRCCSD[T] & MR$-$LDSRG(2)  \\
\midrule
Be$_2$      & (4,8)         &   2.448&           0.949&           2.230&           2.228&           1.527 \\
BeH$_2$     & (4,6)         &   0.461&           0.181&           1.597&           1.551&           2.563 \\
BH$_3$      & (6,7)         &   1.116&           0.281&           1.454&           1.408&           3.769 \\
BH		    & (4,6)         &   1.070&           0.307&           1.052&           1.102&           1.858 \\
C$_2$H$_2$	& (8,8)         &   4.801&          $-$1.260&           0.645&           0.076&           4.340 \\
C$_2$		& (6,6)         &  20.439&          $-$2.362&           4.129&           4.051&           4.077 \\
CH$_2$O		& (8,8)         &   5.059&          $-$0.869&          $-$0.562&          $-$0.060&           5.094 \\
CH$_4$		& (8,8)         &   0.698&          $-$0.909&          $-$0.708&          $-$0.662&           4.025 \\
CO		    & (6,6)         &   7.501&          $-$0.805&          $-$0.318&           0.277&           4.512 \\
F$_2$		& (2,2)         &   5.143&          $-$0.321&           0.829&          $-$0.371&           3.100 \\
H$_2$O$_2$	& (2,2)         &   4.425&          $-$0.453&           4.033&           3.020&           1.898 \\
H$_2$O		& (4,4)         &   0.601&          $-$0.395&          $-$3.390&          $-$3.757&           4.982 \\
H$_2$		& (2,2)         &   0.000&           0.000&           0.180&           0.094&           2.085 \\
HCN		    & (6,6)         &   6.331&          $-$0.656&           0.344&          $-$0.198&           4.583 \\
HF			&  (2,2)        &   0.150&          $-$0.477&          $-$1.836&          $-$2.230&           1.843 \\
HNC			& (10,9)        &   6.920&          $-$1.118&           0.817&           0.726&           4.996 \\
HNO			& (6,6)         &   7.252&          $-$0.229&          $-$0.468&          $-$0.843&           4.549 \\
HOF			& (2,3)         &   4.650&          $-$0.491&           1.947&           1.557&           1.544 \\
Li$_2$		& (2,5)         &  $-$0.689&          $-$0.734&          $-$0.713&          $-$0.705&          $-$0.707 \\
LiH			& (2,3)         &   0.011&           0.001&           0.601&           0.604&           1.759 \\
N$_2$H$_2$	& (6,6)         &   5.935&          $-$0.795&           2.868&           1.838&           6.881 \\
N$_2$		& (10,8)        &   7.406&          $-$0.177&          $-$2.491&          $-$3.145&           2.800 \\
NH$_3$		& (4,4)         &   0.815&          $-$0.621&          $-$0.266&          $-$0.053&           1.549 \\
\midrule
MAE        &    &        4.083&           0.626&           1.456&           1.328&           3.263\\
MAX        &    &       20.439&           2.362&           4.129&           4.051&           6.881\\
\bottomrule
\end{tabular}}
\caption{Single point energy differences of single and multireference CC methods with respect to FCI in milliHartree in the 6-31G basis set. Additionally, we provide mean absolute errors (MAE) and maximum errors (MAX) in milliHartree for every method. The active spaces were selected starting from the full valence active space based on occupation numbers. FCI results and structures are taken from Ref.~\citenum{Evangelista2020_ldsrg3}.}
\label{tab:mol_set}
\end{table}

We first analyze deviations from FCI for 23 molecules selected from the test set employed by Li and Evangelista\cite{Evangelista2020_ldsrg3}. The active spaces were selected by first performing full-valence CASSCF calculations, and subsequently, we eliminated orbitals with natural occupation numbers close to zero or two. The results are presented in Table \ref{tab:mol_set}, with FCI energies and structures taken from Ref.~\citenum{Evangelista2020_ldsrg3}. 

Table \ref{tab:mol_set} shows that CCSD(T) is the most accurate method with an MAE of less than 1~milliHartree, while CCSD is the most inaccurate method with a substantial maximum error of 20.439~milliHartree for C$_2$.
Interestingly, ric-MRCCSD and ric-MRCCSD[T] perform very similarly to one another.
The triples correction yields a slightly lower MAE and maximum error.
The accuracy of both methods is in between CCSD and CCSD(T).
Also, both ric-MRCC methods perform better than MR-LDSRG(2).

\begin{table}
    \centering
\resizebox{\textwidth}{!}{%
\begin{tabular}{l r r r r r r}
\toprule
          & CCSD & CCSD(T) & NEVPT2      &   ric-MRCCSD      & ric-MRCCSD[T]  & Experiment  \\
\midrule           
cc-pVTZ &&&&&\\
\midrule           
        $r_e$ / \AA                &     1.094    &      1.101      &          1.103  &      1.100   &         1.101     &   1.098   \\
        $\omega_e$ / cm$^{-1}$     &    2443      &     2365        &         2341    &    2364   &      2361      &   2358 \\ 
        $\omega_e x_e$ / cm$^{-1}$ &    13.91     &     14.56       &        11.18    &     14.80    &          14.98   &   14.32   \\
\midrule         
cc-pVQZ &&&&&\\
\midrule           
        $r_e$ / \AA                &    1.091 & 1.098 & 1.100      & 1.098 & 1.098 & 1.098 \\ 
        $\omega_e$ / cm$^{-1}$     &    2448  & 2368  & 2353      &  2372 & 2369 & 2358 \\ 
        $\omega_e x_e$ / cm$^{-1}$ &    13.91 & 14.57 & 19.86      & 14.78 & 14.88 & 14.32 \\  
\bottomrule
    \end{tabular}}
    \caption{Spectroscopic constants for N$_2$ evaluated with different SRCC, MRCC, and MRPT methods with two larger basis sets. $r_e$ corresponds to the equilibrium bond length, $\omega_e$ is the harmonic constant and $\omega_e x_e$ is the anharmonic constant.
    The experimental values are taken from Ref.~\citenum{NIST2024}.}
    \label{tab:n2_constants}
\end{table}

Due to the renormalization, neither DSRG nor ric-MRCC methods are guaranteed to converge to the FCI energy at finite values of $s$.
However, relative errors are more important.
Consequently, we continue by comparing spectroscopic constants of N$_2$ -- specifically, the equilibrium bond length ($r_e$), the harmonic constant ($\omega_e$), and the anharmonic constant\cite{herzberg1950spectra} ($\omega_e x_e$) -- with results from SRCC calculations and experimental data \cite{NIST2024}.
We used a fourth-order Taylor series centered around 1.1038~{\AA} with intervals of 0.025~{\AA} to estimate $r_e$ and the derivatives.
The constants, $\omega_e$ and $\omega_e x_e$ were evaluated following Ref.~\citenum{Crawford1996_ComparisonTwo}.

We performed these calculations using two different basis sets, cc-pVTZ and cc-pVQZ, and we summarize the results in Table \ref{tab:n2_constants}.
For the cc-pVTZ basis set, the $r_e$ values are all within 0.005 Å of the experimental data.
However, there is a significant variation in the harmonic constants.
For example, the CCSD harmonic constant differs by 85 cm$^{-1}$ from the experimental value, whereas NEVPT2 shows a deviation of 17~cm$^{-1}$.
The discrepancy of the CCSD(T) harmonic constant is much smaller with 7~cm$^{-1}$.
In terms of the anharmonic constants, NEVPT2 shows a larger deviation of 3.28~cm$^{-1}$ compared to 0.24~cm$^{-1}$ for CCSD(T), indicating that NEVPT2 is less accurate farther away from the equilibrium structure.
The results of the ric-MRCC methods are closely aligned with those of CCSD(T).
The same trends are observed for the cc-pVQZ basis set.

In particular, for this larger basis set, the difference between ric-MRCCSD[T] and CCSD(T) is less than 0.001~{\AA} for $r_e$, about 1~cm$^{-1}$ for the harmonic frequency, and 0.31~cm$^{-1}$ for the anharmonic constant.
We can conclude that for the spectroscopic constants of N$_2$, the ric-MRCC methods are more accurate than CCSD and NEVPT2 and match the accuracy of CCSD(T).

\subsection{Chromium dimer}

Finally, we analyze the PES of the chromium dimer with NEVPT2, ric-MRCCSD, and ric-MRCCSD[T] with a CAS(12,12) and compare these results to the best DMRG estimates available within the cc-pVDZ-DK basis set\cite{Larsson2022_ChromiumDimer}.
The chromium dimer is one of the most notorious test molecules for multireference methods, as extensively discussed in Ref.\citenum{Larsson2022_ChromiumDimer}.

\begin{figure}
    \centering
    \includegraphics[width=0.7\textwidth]{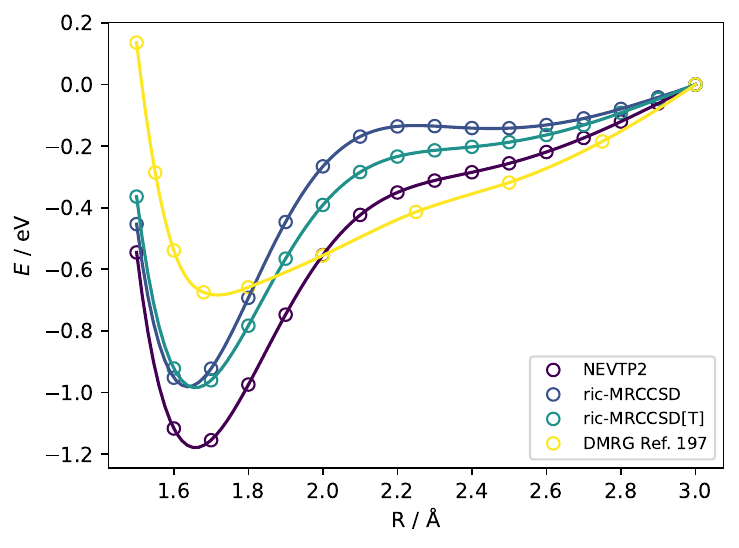}
    \caption{Potential energy curve of the chrmomium dimer in the cc-pVDZ-DK basis set with a CAS(12,12). The curves are aligned to coincide at 0~eV at 3~{\AA}. Shown are results for the best DMRG FCI estimate\cite{Larsson2022_ChromiumDimer}, ric-MRCCSD, ric-MRCCSD[T], and NEVPT2.}
    \label{fig:cr2}
\end{figure}

We note that the limitations of pilot implementation are forcing us to employ the small cc-pVDZ basis set with a minimal CAS.
Kurashige et al.\ showed in Ref.~\citenum{Kurashige2011_SecondorderPerturbation} that most of the error of CASPT2 in a CAS(12,12) can be ascribed to the missing 4$p$ and 4$d$ orbitals in the active space.
It was demonstrated that CASPT2 with a CAS(12,12) qualitatively reproduces the features of the PES, whereas employing a CAS(12,28) yields semi-quantitative results.

Hence, the goal is to qualitatively capture important features of the PES, such as the equilibrium bond length and the distinctive shelf region -- features that are not reproduced adequately by symmetry-broken CCSD(T).\cite{Bauschlicher1994_Cr2Revisited}
The curves are presented in Fig.~\ref{fig:cr2}, and for the sake of comparability, we aligned all results to the same zero-reference energy at a distance of 3 {\AA}. 
As previously reported in the literature\cite{Angeli2001_NelectronValence}, NEVPT2 -- like CASPT2\cite{Kurashige2011_SecondorderPerturbation,Battaglia2022_RegularizedCASPT2} -- qualitatively reproduces the PES with a CAS(12,12), though it lacks quantitative accuracy. ric-MRCCSD improves over NEVPT2 but introduces an unphysical local maximum in the shelf region, a phenomenon previously observed with NEVPT2 in a different basis set\cite{Angeli2001_NelectronValence}. Adding the perturbative triples correction improves the PES shape further and removes the unphysical maximum. 
Additionally, with all methods, the minimum of the PES is qualitatively close to the DMRG minimum. 

In summary, the ric-MRCC methods surpass NEVPT2 in accuracy. However, due to the constraints of the active space size, they fail to achieve quantitative agreement with the DMRG PES.

\section{Conclusions}

In this work, we have built on developments of ic-MRCC ,\cite{Evangelista2011_OrbitalinvariantInternally,Hanauer2011_PilotApplications,Hanauer2012_PerturbativeTreatment,Datta2011_StatespecificPartially} SRG ,\cite{Tsukiyama2011_InMediumSimilarity,Hergert2016_InmediumSimilarity,Hergert2016_InMediumSimilaritya} and DSRG -- especially the iterative multireference generalization, MR-LDSRG(2)\cite{Evangelista2014_DSRG1,Evangelista2015_MRDSRG,Evangelista2016_RelaxedDSRG} -- to combine all of these approaches in the ric-MRCC method. 
For our development, we relied on the generalized normal ordering,\cite{Mukherjee1997_NormalOrdering,Kutzelnigg1997_NormalOrdering} and automatic code generation\cite{Evangelista2022_Wicked} to derive the complex equations of internally contracted methods. 
A problem of ic-MRCC methods is that they become numerically unstable when the active space contains almost redundant orbitals.\cite{Datta2011_StatespecificPartially}
To eliminate the numerical instability, MR-LDSRG(2) uses the SRG, which provides a rationale for renormalizing the amplitudes and is governed by the renormalization parameter, $s$. In this ansatz, the many-body formulation of the unitary ic-MRCC\cite{Hoffmann1988_UnitaryMulticonfigurational,Chen2012_OrbitallyInvariant} method is recovered when $s$ tends to infinity. 

The SRG approach is traditionally formulated as a unitary transformation theory, and in this work, we have shown that it can also be extended to nonunitary transformations by introducing an asymmetric Wegner\cite{Wegner1994_FlowequationsHamiltoniansa} generator. This provided the foundation for formulating a nonunitary DSRG theory, i.e., the renormalized ic-MRCC (ric-MRCC) theory. 
The advantage of the nonunitary similarity transformation is the faster convergence of the BCH expansion. 
The DSRG theory relies on the recursive single commutator expansion,\cite{Yanai2006_CanonicalTransformation} which comes with two disadvantages: due to a mismatch in some prefactors, the expansion may introduce substantial errors,\cite{Evangelista2012_ApproximationSimilaritytransformed} and up to ten nested commutators must be evaluated to achieve convergence of the BCH expansion which results in a substantial computational cost. 

Given the faster convergence of the nonunitary transformation, we could introduce a new approximation to the BCH expansion. Evangelista and Gauss\cite{Evangelista2011_OrbitalinvariantInternally} found that already with the double commutator, the BCH expansion of the ic-MRCC method is converged to very high accuracy. However, evaluating the entire double commutator is computationally unfeasible for all but model systems. Therefore, we introduced several approximations to evaluate the commutator, resulting in a computationally feasible yet accurate approximation. 

By including perturbative triples in ric-MRCCSD[T], we achieved very high accuracy compared to FCI, DMRG, and CCSD(T).
We showed that the NPE of ric-MRCCSD[T] is below 3~milliHartree for dissociating several small molecules with notorious multireference character.
Additionally, we compared absolute energies for a set of 23 molecules against FCI.
Here, we found that both ric-MRCCSD and ric-MRCCSD[T] perform significantly better than CCSD and MR-LDSRG(2) and are relatively close to CCSD(T).
However, properties are of more practical relevance than absolute energies.
Hence, we also evaluated spectroscopic constants of the N$_2$ molecule and compared them against SRCC methods and NEVPT2.
Here, we found that both ric-MRCC methods significantly outperform CCSD and NEVPT2 and are on par with CCSD(T).
Finally, we evaluated the potential energy curve of the chromium dimer with NEVPT2 and the ric-MRCC methods in a minimal CAS(12,12) and compared it against the most accurate DMRG estimates\cite{Larsson2022_ChromiumDimer}.
Both ric-MRCC methods are qualitatively more accurate than NEVPT2, and ric-MRCCSD[T] removes an unphysical maximum in the shelf region of the ric-MRCCSD PES.
However, the CAS(12,12) was insufficient to achieve quantitative accuracy with our methods.
Comparing ric-MRCCSD and ric-MRCCSD[T], we can summarize that the latter approach is more accurate when encountering very strong correlation, as in the dissociation of N$_2$ and Cr$_2$, while both methods perform equally well for weak to medium correlation.

In future work, the primary goal should be to develop a more efficient implementation of the ric-MRCC methods -- ideally in the spin-free formulation\cite{Li2021_SpinfreeFormulation}. This will enable one to go beyond model systems and target large active spaces. Given that we were already able to target a CAS with 12 orbitals in the pilot spin orbital-based implementation, we expect that in conjunction with DMRG calculations, active spaces with up to 30 orbitals will be possible.
Employing schemes for approximating the three-body cumulant\cite{Zgid2009_StudyCumulant,Li2023_IntruderfreeCumulanttruncated} or investigating the possibilities of approximating higher-order cumulants introduced by Lechner\cite{lechner2023_icMRCC-Thesis} may facilitate to go even beyond this limit.
We also note that we did not discuss relaxing the reference coefficients,\cite{Evangelista2016_RelaxedDSRG} which should also be a subject of future studies.

In addition, this work paved the way for formulating a nonunitary version of the IM-SRG, enabling us to remove the dependence on the renormalization parameter. Instead of optimizing the amplitude equations, one would solve the SRG flow equations.

\section*{Acknowledgments}
We gratefully acknowledge financial support from the Swiss National Science Foundation (grant no. 200021\_219616).

\providecommand{\latin}[1]{#1}
\makeatletter
\providecommand{\doi}
  {\begingroup\let\do\@makeother\dospecials
  \catcode`\{=1 \catcode`\}=2 \doi@aux}
\providecommand{\doi@aux}[1]{\endgroup\texttt{#1}}
\makeatother
\providecommand*\mcitethebibliography{\thebibliography}
\csname @ifundefined\endcsname{endmcitethebibliography}
  {\let\endmcitethebibliography\endthebibliography}{}

\end{document}